# Deriving the second law of thermodynamics and exploring its boundaries

Yu Qiao[1,2,*]

[1] *Program of Materials Science and Engineering, University of California – San Diego, La Jolla, CA 92093, U.S.A.*

[2] *Department of Structural Engineering, University of California – San Diego, La Jolla, CA 92093-0085, U.S.A.*

[*] *Email: yqiao@ucsd.edu*

**Abstract.** The second law of thermodynamics still lacks a general proof applicable to classical and quantum systems across broad ranges of degrees of nonequilibrium, particle interactions, and time scales. In this paper, we show that from Boltzmann's assumption of equal *a priori* probabilities, the second law naturally emerges via the phase-space continuity equation for the macrostate-level probability $f$ defined over fixed $\Omega$-intervals, where $\Omega$ is the number of microstates. Further analysis suggests a more general assumption underlying the second law: $f$ increases monotonically with $\Omega$, i.e., $\partial f/\partial\Omega > 0$. Based on these findings, the second law can be readily derived for fully chaotic systems. The derivation does not rely on dynamical details, highlighting the statistical nature of entropy increase. For the dilute hard-sphere gas model considered in the H-theorem, the second law may also be established through a simplified approach that requires the Boltzmann equation but does not use the H-function. In contrast, if the positive correlation between $f$ and $\Omega$ breaks down, i.e., $\partial f/\partial\Omega \leq 0$ (e.g., because of local nonchaoticity), the second law of thermodynamics may not apply: entropy can decrease spontaneously with no energetic penalty, and the system can reach an intrinsic-nonequilibrium steady state.

## 1. Introduction

Suppose we want to develop a computer game in which, despite possible nonphysical features (e.g., magical incarnations), the first and second laws of thermodynamics must be strictly enforced. What algorithms can ensure these requirements? Here, the computer game serves as a model of an isolated system.

For a Hamiltonian system, the key to satisfying the first law of thermodynamics is relatively well understood through Noether's theorem [1]: time-translation symmetry. As long as the governing equations are unchanged under a uniform translation of time, energy remains conserved. In contrast, for the second law, because a rigorous proof with well-defined boundaries is still lacking, there has not been a definitive answer.

A central challenge in establishing the second law is to explain why entropy increase is macroscopically irreversible in an isolated system, while the equations of motion (e.g., Newton's laws or the Schrödinger equation) are microscopically reversible. Among the quantitative attempts in this direction, Boltzmann's H-theorem is the most influential [2]. It is based on the hypothesis of molecular chaos: before a collision between two gas particles, the two-body probability density ($\rho_{\mathrm{II}}$) is replaced by the product of the corresponding one-body probability densities ($\rho_{\mathrm{a}}$ and $\rho_{\mathrm{b}}$); that is, $\rho_{\mathrm{II}} = \rho_{\mathrm{a}}\rho_{\mathrm{b}}$. Hence, the probability of microstate evolution is time-asymmetric. From the Boltzmann equation, a one-body entropy $S_{\mathrm{H}} = -k_{\mathrm{B}}H$ is defined, where $k_{\mathrm{B}}$ is the Boltzmann constant, $H = \int \rho_{\mathrm{a}} \ln \rho_{\mathrm{a}} \, \mathrm{d}\Gamma$ is the H-function, and $\mathrm{d}\Gamma$ denotes integration over the phase space. The form of $S_{\mathrm{H}}$ is similar to the Gibbs entropy, and its time evolution is monotonic, i.e., $\dot{S}_{\mathrm{H}} \geq 0$, where the dot notation denotes the time derivative.

However, it is well known that the H-theorem has limitations. First, the H-theorem has a relatively narrow scope [3]. It is applicable to classical ideal gases within the Boltzmann-Grad limit [4], systems with short-range interactions, and certain weakly correlated particles. It cannot be directly applied to condensed matter, short-time dynamics, long-range interactions, etc. Second, as suggested by Loschmidt's paradox and the Poincaré recurrence theorem, entropy increase is statistical rather than dynamical [5,6]. Entropy-decreasing evolution of individual microstates is suppressed by the large phase-space volume ratios and the high sensitivity to perturbations. Third, $\rho_{\mathrm{II}} = \rho_{\mathrm{a}}\rho_{\mathrm{b}}$ is inapplicable to many far-from-equilibrium states [2]. Fourth, the H-theorem does not guarantee equilibration [7-9]. Although $\dot{S}_{\mathrm{H}} \geq 0$, the mechanism by which $S_{\mathrm{H}}$ converges to its maximum is not fully clear. Fifth, $\dot{S}_{\mathrm{H}} \geq 0$ is not the second law of thermodynamics, since $S_{\mathrm{H}}$ is different from the Gibbs entropy [5,10]. The increase in $S_{\mathrm{H}}$ is related to the erased correlations, as the hypothesis of molecular chaos repeatedly re-imposes randomness. Sixth, as a direct motivation for the present study, the H-theorem does not account for the intrinsic-nonequilibrium phenomena reported in [11-14]. For instance, what would be the consequence if particle-particle collisions are not extensive in a local region? Moreover, the H-theorem does not directly address quantum systems. Interestingly, recent research has shown that a bound state inherently satisfies the second law [14]; it is the classical-mechanical and semiclassical systems (e.g., a scattering state with unquantized energy and localized wave packets) that need further investigation.

The concept behind the H-theorem is quite intuitive: because the microscopic equations of motion (EOM) determine the dynamical details, one might expect that the macroscopic properties (e.g., the second law) should also follow from them. Yet, as an emergent phenomenon, the second law may not depend on the specific microscopic dynamics. For instance, the ideal-gas law can be derived without using Newton's laws [14].

Sections 2 and 3 below present a general derivation of the second law of thermodynamics that is independent of microscopic dynamics, the degree of nonequilibrium, the range of correlations, and the time scale, thereby circumventing the aforementioned limitations of the H-theorem. It places the second law on a transparent footing, based on which we then examine the key assumptions, their boundaries, and possible extensions beyond the boundaries (Section 4). Broader discussions are given in Section 5.

## 2. Deriving the second law of thermodynamics

### 2.1 Macrostate-level properties: $f$ and $\Omega$

Unlike the Boltzmann equation formulated for the microstate-level probability [2], our analysis focuses on the macrostate-level probability $f = \sum_{\Omega} \rho_\alpha$, where $\rho_\alpha$ is the probability of the $\alpha$-th microstate, and $\Sigma_\Omega$ denotes summation over the microstates associated with the macrostate or macrostate-like phase-space domain. Hereafter, $f$ is treated as a function of the number of microstates, i.e., the multiplicity $\Omega$. The completeness condition is $\sum_\alpha \rho_\alpha = 1$, where $\Sigma_\alpha$ denotes summation over all microstates under investigation in the phase space.

For an isolated system, the Boltzmann entropy $S = k_B \ln \Omega$ of a macrostate is determined by $\Omega$. Like $f$ and $S$, $\Omega$ is also a macrostate property, a function (denoted by $\Omega_x$) of the macroscopic state variables $\{y_l\}$ ($l = 1,2$ ...) that define the macrostate; that is, $\Omega = \Omega_x(\{y_l\})$. For instance, for a $N$-particle classical ideal gas, $\Omega$ is dependent on the gas volume ($V$) and the probability density distribution of particle momentum ($f_T$). In other words, $\{y_l\}$ includes $\{N, V, f_T\}$ and $\Omega = \Omega_x(N, V, f_T)$. According to the Sackur-Tetrode equation [2], at thermal equilibrium when $f_T$ is the Maxwell-Boltzmann distribution ($f_{\text{MB}}$), $\Omega_{\text{x}}(N,V) = (V/N)^N\left[4\pi e^{5/3}mU/(3h^2N)\right]^{3N/2}$, where $m$ is the particle mass, $U$ is the internal energy, and $h$ is the Planck constant. When the gas body fills the gas container ($V_0$), $\Omega$ reaches the maximum value $\Omega_{\text{max}}$. The minimum $\Omega$ ($\Omega_{\text{min}}$) depends on the nature of the particles.

2.2 Funnel-shaped phase space

Figure 1(a) depicts the phase space of an isolated system. The funnel-shaped volume contains all accessible microstates between $\Omega_{\text{min}}$ and $\Omega_{\text{max}}$. The microstates are arranged along the $\Omega$-axis. From bottom to top, $\Omega$ increases.

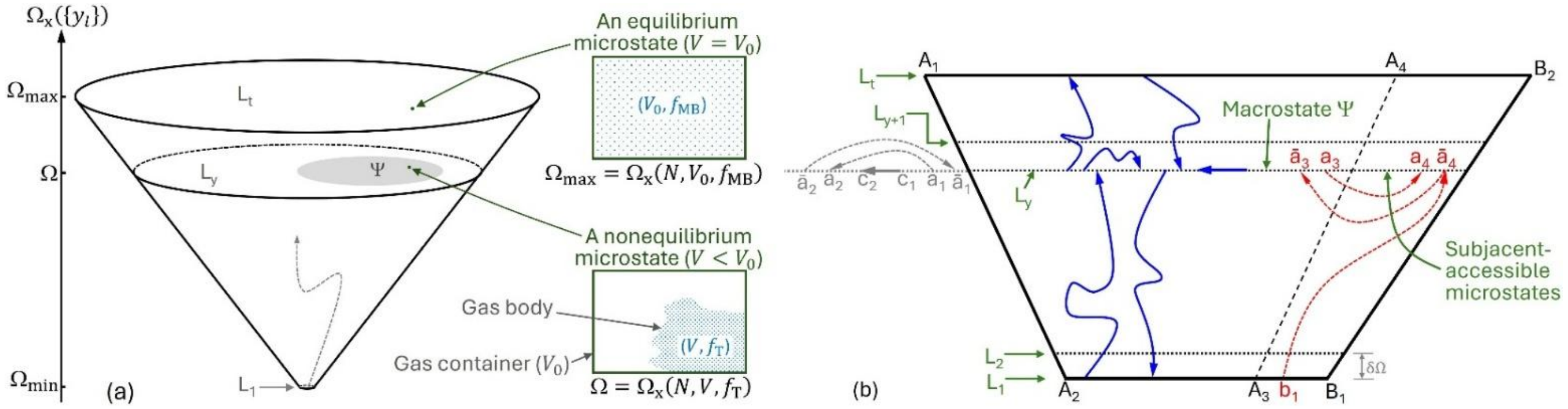

**Figure 1. (a)** Schematic of the funnel-shaped phase space of an isolated system. Each horizontal "layer" (hyperplane) represents a certain value of $\Omega$, i.e., the subset of microstates that belong to the macrostates satisfying $\Omega_{\text{x}}(\{y_l\}) = \Omega$. The layer size visualizes the number of microstates in the subset. From bottom to top, $\Omega$ increases. The grey circle in hyperplane $L_y$ indicates a macrostate ($\Psi$). The grey dotted curve depicts a possible microstate evolution trajectory, starting from the bottom of the phase space ($L_1$). The insets depict two representative states of a classical ideal gas: the upper inset shows an equilibrium state, in which the gas fills the container; the lower inset shows a nonequilibrium state, in which the gas does not fill the container. **(b)** Side view of the phase-space subspace ($A_1A_2B_1B_2$) constructed for macrostate $\Psi$. It consists of two sections: $A_1A_2A_3A_4$ and $B_1B_2A_4A_3$. Within hyperplane $L_y$, the horizontal slice (cross section) of $A_1A_2A_3A_4$ represents the macrostate ($\Psi$), and the horizontal slice of $B_1B_2A_4A_3$ includes all subjacent-accessible microstates. The solid and dashed curves represent possible evolution paths of microstates; $a_1$ is a microstate in $\Psi$ that first evolves upward and then returns to $L_y$ from above as another microstate outside $\Psi$ ($a_2$); $a_3$ is a microstate in $\Psi$ that first evolves downward and then returns to $L_y$ from below as another microstate outside $\Psi$ ($a_4$); $c_1$ evolves within $L_y$ into $c_2$, crossing from inside $\Psi$ to outside $\Psi$; $\bar{a}_1 - \bar{a}_4$ are the time-reversed microstates of $a_1 - a_4$, respectively; time-reversed microstates of $c_1$ and $c_2$ are not shown. Excluding $a_1$, $\bar{a}_1$, and $c_1$ from the phase-space subspace ($A_1A_2B_1B_2$) is conservative for the slice-by-slice analysis in Sections 2.5-2.6.

A horizontal "layer" (hyperplane) corresponds to a certain value of $\Omega$, representing the subset of microstates that belong to the macrostates satisfying $\Omega_{\text{x}}(\{y_l\}) = \Omega$. The layer size visualizes the number of microstates in the subset. In each layer, there may be multiple macrostates. The grey circle in layer $L_y$ represents a macrostate ($\Psi$), i.e., all microstates belonging to $\Psi$. At the top of the phase space ($L_t$), the system is in equilibrium and entropy is maximized at $S_{eq} = k_B \ln \Omega_{max}$. The insets depict two representative states of an ideal gas. The upper inset shows an equilibrium state with $\Omega_x(N, V_0, f_{MB}) = \Omega_{max}$, in which the gas body fills the gas container (i.e., $V = V_0$). The lower inset shows a nonequilibrium state with $\Omega_x(N, V, f_T) < \Omega_{max}$, in which the gas body is concentrated in a corner of the container (i.e., $V < V_0$).

Such a phase-space configuration may be constructed by tracing microstate evolutions. The grey dashed curve in Figure 1(a) illustrates a microstate evolution trajectory over time $t$, starting from the bottom hyperplane $L_1$ where $\Omega_x(\{y_l\}) = \Omega_{min}$. If a microstate evolution trajectory moves upward or remains at the same level of $\Omega$, the $\Omega$-axis is aligned with time $t$. Later microstates are placed in the same or a higher horizontal layer, depending on $\Omega_x(\{y_l\})$. If the evolution trajectory moves downward, forming the funnel requires rearranging the microstates along the $\Omega$-axis from their chronological order. Nevertheless, the positions in the funnel and the temporal sequence are bijectively related, and the rearrangement does not alter the topological structure. The funnel can be completed by tracing all possible microstate evolution paths.

A continuous phase space consists of infinitely many points, rendering the Boltzmann entropy $S = k_B \ln \Omega$ ill-defined. A common resolution is to assign a finite phase-space volume $\delta\Gamma$ to each microstate, so that $\Omega$ is countable. For instance, in calculating ideal-gas entropy, the Sackur-Tetrode equation takes $\delta\Gamma = h^{3N}$ [2]. This discretization is equivalent to fixing the additive constant of entropy and ensures consistency with the semiclassical limit of quantum mechanics. An alternative approach for a non-discretized phase space is to use an arbitrarily fine phase-space resolution. As microstates evolve between the bottom hyperplane ($L_1$) and the top hyperplane ($L_t$) in the funnel-shaped phase space (not including the evolution paths within $L_1$ and $L_t$), whenever an evolution path branches, instead of re-averaging the microstate probability density ($\rho$) in a Boltzmann-type coarse-grained description [4,5], each branch, together with all microstates that evolve into it, is treated as a new independent path. The two methods yield convergent results if $\delta\Gamma$ is smaller than the branch size. For simplicity, the discussion below adopts the former framework of discretized phase space.

In Figure 1(a), the "thickness" of each layer is fixed (denoted by $\delta\Omega$), which reflects the statistical nature of macrostates and can be viewed as the resolution of $\Omega$. The scale of $\delta\Omega$ should be larger than that of $\delta\Gamma$, so that the probability distribution of microstates ($\rho$) can be defined in macrostates. Effectively, the multiplicity axis is partitioned into intervals of equal width, without averaging $\rho$ over phase-space cells.

For the microstate evolution processes under investigation, the time resolution $\delta t$ is sufficiently fine, so that $\delta t < \delta\Omega/\dot{\Omega}_{\mathrm{m}}$, where $\dot{\Omega}_{\mathrm{m}}$ is the maximum rate of change of $\Omega$ as the system evolves. In the funnel-shaped phase space, during each timestep $\delta t$, a microstate evolves either within the same constant-$\Omega$ hyperplane (layer) or into an adjacent hyperplane (layer).

## 2.3 Phase-space subspace for a macrostate

Figure 1(b) depicts the phase-space subspace ($A_1A_2B_1B_2$) constructed for macrostate $\Psi$. It comprises two sections: $A_1A_2A_3A_4$ and $B_1B_2A_4A_3$. Within the constant-$\Omega$ hyperplane $L_y$, the horizontal slice (cross section) of $A_1A_2A_3A_4$ represents $\Psi$. Section $A_1A_2A_3A_4$ contains all microstates along the evolution trajectories that, between the top ($L_t$) and bottom ($L_1$) of the phase space, either originate from or terminate in $\Psi$, without revisiting $L_y$ after leaving it, as shown by the blue solid curves. It also contains all evolution paths between two microstates of $\Psi$. The segments of evolution paths inside $L_t$ or $L_1$ are not included.

The slice of $A_1A_2A_3A_4$ is only a subset of $\Psi$. There are three other subsets of microstates in $\Psi$. From the first subset, a microstate of $\Psi$ (e.g., $a_1$) evolves upward and then returns to $L_y$ from above as a microstate ($a_2$) of a different macrostate outside $\Psi$; it will be termed superjacent-accessible. From the second subset, a microstate of $\Psi$ (e.g., $c_1$) evolves within $L_y$ into a microstate ($c_2$) outside $\Psi$; it will be termed coplanar-accessible. From the third subset, a microstate of $\Psi$ (e.g., $a_3$) evolves downward and then returns to $L_y$ from below as a microstate ($a_4$) outside $\Psi$; it will be termed subjacent-accessible. For reasons discussed later, the first two subsets (including the time-reversed microstates) are excluded from the phase-space subspace $A_1A_2B_1B_2$. For the third subset, in accordance with micro-reversibility [15], for every outward evolution trajectory (e.g., $a_3a_4$) there is a time-reversed inward trajectory $\bar{a}_4\bar{a}_3$, i.e., a microstate ($\bar{a}_4$) in $L_y$ outside $\Psi$ leaves $L_y$ and evolves into a microstate of $\Psi$ ($\bar{a}_3$) from below; $\bar{a}_3$ and $\bar{a}_4$ are the time-reversed microstates of $a_3$ and $a_4$, respectively. Section $B_1B_2A_4A_3$ contains all microstates that either can evolve into or be reached from a subjacent-accessible microstate or its time-reversed counterpart (e.g., $b_1\bar{a}_4$) without revisiting $L_y$ after leaving it. If a subjacent-accessible microstate outside $\Psi$ has a subjacent-accessible microstate, then the new microstate and its time-reversed counterpart, along with all microstates that can either evolve into or be reached from them (without revisiting $L_y$ after leaving it), are also included in $B_1B_2A_4A_3$. This process continues until all subjacent-accessible possibilities are exhausted.

The phase-space subspace ($A_1A_2B_1B_2$) is designed to be closed, i.e., no microstate in it evolves across its lateral boundary. Microstate trajectories within the top ($L_t$) and bottom ($L_1$) hyperplanes do not affect the derivation of the second law.

## 2.4 Continuity equation of $f$ in Figure 1(b)

In Figure 1(b), the probability of a horizontal slice of the phase-space subspace $A_1A_2B_1B_2$ is defined as $f = \sum_S \rho_\alpha$, with $\Sigma_S$ denoting summation over the microstates under investigation in the slice. Along the $\Omega$-axis, with a fixed $\delta\Omega$, the continuity equation for $f$ is [16]

$$\frac{\partial f}{\partial t} = -\dot{\Omega}\frac{\partial f}{\partial \Omega} - f\frac{\partial \dot{\Omega}}{\partial \Omega}, \tag{1}$$

where $\dot{\Omega}$ should be viewed as the expected rate of change in the multiplicity of the macrostates to which the microstates in a slice belong. The partial derivative with respect to $\Omega$ may equivalently be understood as $\partial/\partial\Omega_x$, particularly for the setting considered in Sections 2.5-2.6 below; the notation $\partial/\partial\Omega$ is used for convenience when discussing the physical meanings.

Equation (1) does not directly govern the microstate probability ($\rho$) and is unrelated to the coarse-graining of $\rho$. Changes in $f$ along "height" are associated with the compressibility of the probability flow of $f$, i.e., $\partial\dot{\Omega}/\partial\Omega \neq 0$ (see Section 5.4). The specific form of $\partial\dot{\Omega}/\partial\Omega$ depends on $\Omega_x(\{y_l\})$ and microscopic dynamics.

## 2.5 Derivation of the second law of thermodynamics

Without additional assumptions, Equation (1) by itself offers no particularly useful information about the evolution of $f$ or $\Omega$. Let $\mathcal{B}$ denote Boltzmann's assumption of equal *a priori* probabilities [5,17,18]: microstates with the same values of the conserved quantities have equal probability, irrespective of the macrostates to which they belong. For convenience of comparison with common formulations of statistical mechanics [e.g., 18], in the following discussion we consider conservation of energy, under which $\mathcal{B}$ is written as

$$\rho = P_{\mathrm{E}}(\epsilon), \tag{2}$$

with $\epsilon$ being the energy level, and $P_{\mathrm{E}}$ a certain function of $\epsilon$. When necessary, other conserved quantities can be treated in the same manner as $\epsilon$, without affecting the general procedure for deriving the second law.

In an isolated system, all accessible microstates have the same energy and therefore, in accordance with $\mathcal{B}$ (Equation 2), have the same $\rho_\alpha$ [18]: $\rho_0 = 1/\Omega_{\mathrm{t}}$, where $\Omega_{\mathrm{t}}$ is the total number of the microstates under investigation. Hence, for a horizontal slice in Figure 1(b), $f = \sum_{\mathrm{S}} \rho_\alpha$ is simplified as

$$f = \rho_0 \Omega, \tag{3}$$

In Equation (3), $\Omega$ is the number of the microstates under investigation in the slice.

Set the initial condition of Equation (1) as $f = \sum_{\mathrm{S}} \rho_\alpha$ for all microstates at every slice from $\mathrm{L_1}$ to the slice above $\mathrm{L_y}$ (denoted by $\mathrm{L_{y+1}}$). That is, initially $f = \rho_0 \Omega_{\mathrm{x}}$ at each $\Omega_{\mathrm{x}}$ from $\mathrm{L_1}$ to $\mathrm{L_{y+1}}$, with $\Omega_{\mathrm{t}}$ being the number of all microstates from $\mathrm{L_1}$ to $\mathrm{L_{y+1}}$; $f = 0$ above $\mathrm{L_{y+1}}$. Within a short time interval $\delta t$, the slices beyond $\mathrm{L_{y+1}}$ do not affect the probability flow of $f$ at or below $\mathrm{L_y}$.

At $t = 0$, from $\mathrm{L_1}$ to $\mathrm{L_{y+1}}$, with the given initial $f$ and a constant $\rho_0$, the right-hand side of Equation (1) can be rewritten as $-\rho_0 \dot{\Omega} - f\left(\partial \dot{\Omega}/\partial \Omega\right)$, and Equation (1) becomes

$$\dot{\Omega} = -\frac{1}{\rho_0}\left(\frac{\partial f}{\partial t} + f\frac{\partial \dot{\Omega}}{\partial \Omega}\right). \tag{4}$$

We then allow all microstates from $\mathrm{L_1}$ to $\mathrm{L_{y+1}}$ to evolve over a short time interval $\delta t$, each according to its microscopic dynamics.

First, notice that at the lowest horizontal slice $\mathrm{L_1}$ in Figure 1(b), $\dot{\Omega} \geq 0$. At the bottom of the funnel, a microstate in $\mathrm{L_1}$ either evolves into a higher slice or stays in $\mathrm{L_1}$. After $\delta t$, the expectation value of $\Omega_{\mathrm{x}}(\{y_l\})$ associated with the microstates initially in $\mathrm{L_1}$ cannot decrease.

Next, we analyze $\dot{\Omega}$ at the second-lowest slice ($\mathrm{L_2}$). Because initially (at $t = 0$) $f$ accounts for all microstates in the slice, after $\delta t$, $f$ cannot increase, i.e., $\partial f/\partial t \leq 0$ at $\mathrm{L_2}$. For $\partial \dot{\Omega}/\partial \Omega$ at $\mathrm{L_2}$, there are two possible scenarios: $\partial \dot{\Omega}/\partial \Omega \leq 0$ or $\partial \dot{\Omega}/\partial \Omega > 0$. If $\partial \dot{\Omega}/\partial \Omega \leq 0$, Equation (4) suggests that $\dot{\Omega} \geq 0$. If $\partial \dot{\Omega}/\partial \Omega > 0$, at the slice immediately below $\mathrm{L_2}$ (i.e., $\mathrm{L_1}$), with a finite $\left|\partial^2 \dot{\Omega}/\partial \Omega^2\right|$, since $\mathrm{L_2}$ can be arbitrarily close to $\mathrm{L_1}$ (i.e., the layer thickness $\delta\Omega$ can be smaller than

the ratio of $\partial\dot{\Omega}/\partial\Omega$ to $|\partial^2\dot{\Omega}/\partial\Omega^2|$), $\partial\dot{\Omega}/\partial\Omega$ is nonnegative; as a result, with $\dot{\Omega} \geq 0$ at the lower slice ($L_1$), we also have $\dot{\Omega} \geq 0$ at the upper slice ($L_2$).

For the derivation of $\dot{\Omega} \geq 0$ at $L_2$, a setup with a large $\delta\Omega$ is irrelevant, since it neither supports nor disproves $\dot{\Omega} \geq 0$ and yields no useful result. The assumption of a finite $|\partial^2\dot{\Omega}/\partial\Omega^2|$ is reasonable for a finite $\Omega_{max}$ in Figure 1(b) (see Section 2.2), a nonsingular Hamiltonian, and well-behaved EOM (e.g., Newton's laws or the Schrödinger equation) within the framework of the first law of thermodynamics (time-translation symmetry [1]).

An analysis similar to that of $L_2$ can be repeated slice by slice in Figure 1(b), starting from $L_2$ and proceeding upward to $L_y$. Regardless of whether $\partial\dot{\Omega}/\partial\Omega \leq 0$ or $\partial\dot{\Omega}/\partial\Omega > 0$ at each slice, at $L_y$ we have

$$\dot{\Omega} \geq 0. \tag{5}$$

The slice-by-slice procedure would be inconclusive if it were performed from the top layer ($L_t$) downward, due to the mismatch between the direction and the consequence of $\partial\dot{\Omega}/\partial\Omega > 0$.

The coplanar-accessible microstates never leave $L_y$ and make no contribution to $\dot{\Omega}$ at $L_y$. Owing to micro-reversibility [15], every evolution path of superjacent-accessible microstates (e.g., $a_1a_2$) has a time-reversed counterpart (e.g., $\bar{a}_2\bar{a}_1$), with $\bar{a}_1$ and $\bar{a}_2$ being the time-reversed microstates of $a_1$ and $a_2$, respectively. Since all $a_1$-type microstates evolve upward, the overall effect reinforces $\dot{\Omega} \geq 0$, regardless of the evolution directions of the $\bar{a}_1$-type microstates. That is, excluding coplanar- and superjacent-accessible microstates (e.g., $a_1$, $\bar{a}_1$, and $c_1$) from the phase-space subspace is a conservative choice. For the subjacent-accessible subset, because all $\bar{a}_4$-type microstates evolve downward, regardless of the evolution directions of the $a_4$-type microstates, the overall effect of the slice of $B_1B_2A_4A_3$ tends to reduce $\dot{\Omega}$. Because $\dot{\Omega} \geq 0$ for the entire slice of $A_1A_2B_1B_2$, Inequality (5) must also be true for the slice of $A_1A_2A_3A_4$, i.e., macrostate $\Psi$.

In Figure 1(a), for every macrostate from $L_2$ to the layer immediately below $L_t$ (denoted by $L_{t-1}$), a phase-space subspace similar to Figure 1(b) can be constructed, and the same slice-by-slice analysis can be performed. Since $S = k_B \ln \Omega$, the result $\dot{\Omega} \geq 0$ proves the second law

$$\dot{S} \geq 0. \tag{6}$$

Inequality (6) is statistical in nature, in which $S$ is the Boltzmann entropy and is a macrostate property. It reflects the expected behavior of macrostates, taking into consideration possible entropy-decreasing evolution of individual microstates.

Although Inequalities (5) and (6) are obtained by applying a specific initial condition to Equation (1) over a small time interval $\delta t$, $\dot{\Omega}$ is ultimately governed by the microscopic EOM rather than by the probability flow of $f$. Since all microstates of macrostate Ψ have been taken into account and the EOM are invariant under time translation, the conclusion of $\dot{\Omega} \geq 0$ and $\dot{S} \geq 0$ (for Ψ) can be extended to arbitrary times and settings. In essence, in Figure 1(b), $\dot{\Omega}$ is an instantaneous property determined by the microscopic dynamics at local slices, whereas the multiplicity-based definition of $f$ and its particular initial condition are merely devices used to reveal the sign of $\dot{\Omega}$ at $\mathrm{L_y}$.

If the macrostate (Ψ) lies in the top layer $\mathrm{L_t}$, the derivative $\partial/\partial\Omega$ is ill-defined locally, such that the above analysis may not be applicable and a negative $\dot{\Omega}$ or $\dot{S}$ is possible at $\mathrm{L_t}$. However, because $\dot{\Omega}$ and $\dot{S}$ are nonnegative from $\mathrm{L_1}$ to $\mathrm{L_{t-1}}$, the negative $\dot{\Omega}$ and $\dot{S}$ at $\mathrm{L_t}$ reflect fluctuations and do not conflict with the second law.

Through ensemble derivation, the Boltzmann entropy $S$ can be generalized to the Gibbs entropy $S_\mathrm{G}$ [18]. Consider a microcanonical ensemble of a system, in which the total number of microstates $N_\mathrm{E} = \sum_\Omega \hat{n}_\alpha$ is much larger than the number of microstates in the phase space, with $\hat{n}_\alpha$ being the number of the $\alpha$-th microstates in the ensemble. The ensemble-level multiplicity is $\bar{\Omega} = N_\mathrm{E}!/(\prod_\alpha \hat{n}_\alpha!)$. Stirling's approximation leads to $\ln\Omega \approx \ln N_\mathrm{E} - N_\mathrm{E}^{-1}\sum_\Omega \hat{n}_\alpha \ln \hat{n}_\alpha = -\sum_\Omega \rho_\alpha \ln \rho_\alpha$, where $\ln\Omega \triangleq \ln\bar{\Omega}/N_\mathrm{E}$ and $\rho_\alpha \triangleq \hat{n}_\alpha/N_\mathrm{E}$. Thus, $S_\mathrm{G} = k_\mathrm{B}\ln\Omega = -k_\mathrm{B}\sum_\Omega \rho_\alpha \ln\rho_\alpha$.

## 2.6 General assumption underlying the second law

Note that $\mathcal{B}$ is a sufficient but not necessary condition for $\dot{\Omega} \geq 0$. A more general assumption for deriving Inequality (5) is

$$\frac{\partial f}{\partial \Omega} > 0. \tag{7}$$

Equation (1) may be rewritten as $(\partial f/\partial\Omega)\dot{\Omega} = -\partial f/\partial t - f\left(\partial\dot{\Omega}/\partial\Omega\right)$. As in Section 2.5, at $t = 0$, set $f$ to account for all microstates in the slices from $\mathrm{L_1}$ to $\mathrm{L_{y+1}}$. Because $\partial f/\partial\Omega > 0$, over a short time interval $\delta t$, $f$ cannot increase at any of these slices, i.e., $\partial f/\partial t \leq 0$. Thus, again using $\partial f/\partial\Omega > 0$, together with $f \geq 0$, $\dot{\Omega} \geq 0$ and $\dot{S} \geq 0$ can be obtained by the same slice-by-slice approach as for Equation (4), regardless of whether $\partial\dot{\Omega}/\partial\Omega > 0$ or $\partial\dot{\Omega}/\partial\Omega \leq 0$ at each slice. Equation (3) implies that $\mathcal{B}$ is a special case of Inequality (7).

It is worth mentioning that $f$ is defined over phase-space layers of fixed thickness $\delta\Omega$, and $\dot{\Omega} \geq 0$ and $\partial f/\partial\Omega > 0$ do not necessarily suggest that $\dot{f} \geq 0$. According to Equation (1), $\dot{f} = -f(\partial\dot{\Omega}/\partial\Omega)$ [16]. The sign of $\partial\dot{\Omega}/\partial\Omega$ is dominated by microscopic dynamics at the local slice and can be either positive or negative. Further discussion is given in Section 5.2.

## 2.7 Fully chaotic systems obey the second law

In principle, the evolution of a system is governed by its microscopic EOM. Within this framework, the probability of a microstate ($\rho$) can be evaluated once the information specifying the microstate is given, e.g., the energy level ($\epsilon$), the position and momentum of every particle, etc.

In a chaotic system, microstate evolution is highly sensitive to initial conditions and microscopic details are "lost" at the macroscopic level, rendering microstate evolution effectively unpredictable despite deterministic EOM [19]. As a result, the EOM may not explicitly enter the determination of $\rho$, and $\rho$ may carry no dependence on microscopic information beyond the conserved quantities. Accordingly, $\rho$ may be assumed to be constrained only by the conserved quantities (e.g., energy conservation), subject to normalization. In this manuscript, such a condition is referred to as "full chaoticity," and such systems are termed "fully chaotic."

For an isolated fully chaotic system, $\rho_\alpha$ is subject only to the completeness constraint [18]

$$\sum_\alpha \rho_\alpha = 1. \tag{8}$$

Under isothermal conditions, the mean internal energy $U$ imposes another constraint on $\rho_\alpha$ [18]

$$\sum_\alpha \rho_\alpha \epsilon_\alpha = U, \tag{9}$$

where $\epsilon_\alpha$ is the energy level of the $\alpha$-th possible microstate. Equation (8) and (9) are standard in statistical mechanics [e.g., 18]. In either the isolated or isothermal case, $\rho_\alpha$ does not depend directly on variables other than $\epsilon_\alpha$, which justifies $\mathcal{B}$ and, in turn, leads to $\partial f/\partial\Omega > 0$ (Inequality 7). With $\mathcal{B}$ and $\partial f/\partial\Omega > 0$, the second law of thermodynamics (Inequality 6) can be readily derived through the phase-space analysis in Sections 2.5-2.6. Further discussion about $\mathcal{B}$ is presented in Section 5.1.

The assumption of full chaoticity is distinct from the hypothesis of molecular chaos ($\rho_{\mathrm{II}} = \rho_{\mathrm{a}}\rho_{\mathrm{b}}$), as summarized in Table 1. The concept of full chaoticity is rooted in chaos theory [e.g., 19,20], emphasizing the effective decoupling of macroscopic behavior from detailed microscopic dynamics. The hypothesis of molecular chaos does not directly address the high sensitivity to initial conditions, nor can it account for the local-nonchaoticity phenomena in which particle-particle collisions are sparse in certain regions [11-14]. Systems with long-range interactions or far from equilibrium may not satisfy the hypothesis of molecular chaos, yet they may be fully chaotic and satisfy Equations (8) and (9) [18]. The Knudsen-gas models discussed in Section 4 and Appendix 1 satisfy the hypothesis of molecular chaos, yet they are not fully chaotic. In a Knudsen-gas cell, before each particle-wall or particle-particle collision, the microstates of the two interacting bodies are statistically independent; however, as $\rho_\alpha$ contains specific information about microstates, the system is less random than a conventional ideal gas, causing nontrivial energetic consequences.

Essentially, as chaos theory emphasizes extreme sensitivity to initial conditions and the resulting unpredictability of detailed microstate evolution [19], it would be conceptually unsatisfying to attribute $\dot{S} \geq 0$ entirely to statistical independence among particle states. In the framework of full chaoticity, the "fullness" of chaoticity is measured by the extent to which detailed information about microscopic dynamics is erased at the macroscopic level. If the specifics of microstate evolution are completely unpredictable, then $\rho$ has no direct correlation with the microscopic EOM; conversely, if certain information carried by $\rho$ is explicitly determined by the EOM, then the microstate evolution retains some degree of predictability.

Table 1. The key assumptions in the conventional framework

| Assumptions | Remarks |
|---|---|
| Hypothesis of molecular chaos ($\rho_{\text{II}} = \rho_{\text{a}}\rho_{\text{b}}$) | • Not involved in the derivation of the second law in this paper (Sections 2.5-2.6).<br>• Based on the statistical independence of pre-collisional particle states.<br>• Justified for a hard-sphere gas within the Boltzmann-Grad setting [21].<br>• Not always consistent with full chaoticity. Systems with long-range interactions or far from equilibrium may not satisfy $\rho_{\text{II}} = \rho_{\text{a}}\rho_{\text{b}}$ but may be fully chaotic; Section 4 and Appendix 1 present Knudsen-gas models that satisfy $\rho_{\text{II}} = \rho_{\text{a}}\rho_{\text{b}}$ but are not fully chaotic. |
| Assumption of equal *a priori* probabilities (denoted by $\mathcal{B}$) | • A special case of $\partial f/\partial\Omega > 0$ (Inequality 7).<br>• May be justified under ergodicity [17,23] or full chaoticity. For a fully chaotic system, because of the extreme sensitivity to initial conditions, detailed information about microscopic dynamics is "lost" at the macroscopic level, rendering microstate evolution effectively unpredictable despite the deterministic equations of motion (EOM); thus, $\rho_\alpha$ is assumed to have no explicit dependence on the EOM. |

## 3. Simplified derivation with additional assumptions

The derivation of the second law (Inequality 5) may be simplified by introducing more assumptions in addition to $\mathcal{B}$ or $\partial f/\partial\Omega > 0$. The main purpose is to reduce the complexity of the phase-space subspace in Figure 1(b).

### 3.1 Macrostate uniqueness

A possible additional assumption is macrostate uniqueness: the governing equations of macrostate evolution exist and have unique solutions. It may be stated as follows: First, a macrostate can be characterized by a set of macroscopic state variables $\{y_l\}$ $(l = 1,2 \ldots)$. Second, the dynamical evolution of $\{y_l\}$ can be described by a set of macroscopic governing equations, such as the Navier-Stokes-Fourier equations for smooth fluid flows. Third, for given initial and boundary conditions, the macroscopic governing equations of $\{y_l\}$ admit unique solutions. Under these conditions, a macrostate evolves into another macrostate, and may itself be viewed as the outcome of the evolution of other macrostates.

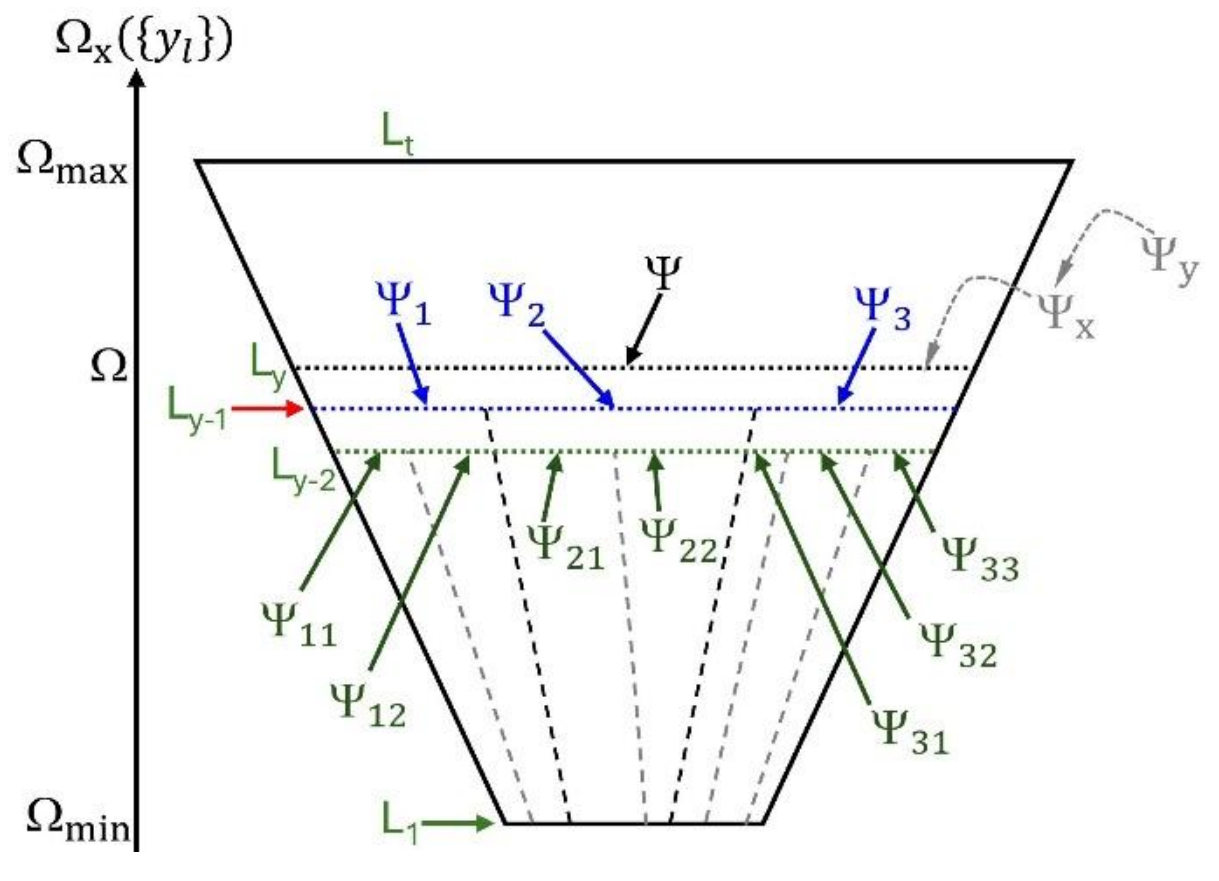

**Figure 2.** Macrostate evolution in the simplified phase-space subspace of $\Psi$.

Figure 2 shows the funnel-shaped phase-space subspace of a macrostate ($\Psi$), with a simpler structure than Figure 1(b). Each horizontal slice (a section of a constant-$\Omega$ hyperplane) corresponds to a certain $\Omega$. Slice $L_y$ represents $\Psi$. The slice below $L_y$ (denoted by $L_{y-1}$) is formed by all macrostates that evolve into $\Psi$, e.g., $\Psi_1$, $\Psi_2$, and $\Psi_3$. The slice below $L_{y-1}$ (denoted by $L_{y-2}$) is formed by all macrostates that evolve into $\Psi_1$, $\Psi_2$, and $\Psi_3$, e.g., $\Psi_{11}$, $\Psi_{12}$, $\Psi_{21}$, $\Psi_{22}$, $\Psi_{31}$, $\Psi_{32}$, and $\Psi_{33}$. This layered structure continues downward to the bottom layer ($L_1$). Macrostates evolving "horizontally" along the layers (constant-$\Omega$ hyperplanes) from outside the phase-space subspace into it do not affect the discussion of the sign of $\dot{\Omega}$.

The slice-by-slice approach can be applied to Figure 2 in the same manner as for Figure 1(b). Since $\dot{\Omega} \geq 0$ at $L_1$, regardless of whether $\partial\dot{\Omega}/\partial\Omega \leq 0$ or $\partial\dot{\Omega}/\partial\Omega > 0$, $\dot{\Omega} \geq 0$ is true for every slice, including $L_y$ (i.e., macrostate $\Psi$). For all macrostates, similar phase-space subspaces may be constructed and $\dot{\Omega} \geq 0$ can be derived.

Note that in Figure 2, no macrostate evolves into $\Psi$ from above. If such a macrostate (denoted by $\Psi_x$) existed, then $\dot{\Omega}$ would have to be negative along a certain portion of its evolution trajectory. This negativity would necessarily arise from other macrostates (denoted by $\Psi_y$) evolving into $\Psi_x$ from above; otherwise, according to the slice-by-slice analysis for the phase-space subspace of $\Psi_x$, a negative $\dot{\Omega}$ would be impossible. In turn, for $\dot{\Omega} < 0$ to occur for $\Psi_y$, yet another set of macrostates would have to evolve into $\Psi_y$ from above. One can therefore trace a chain of macrostates with negative $\dot{\Omega}$ upward until reaching the top layer $L_t$. However, at $L_t$, no macrostates can come from above. Consequently, the chain of negative-$\dot{\Omega}$ macrostates does not exist, and $\dot{\Omega} \geq 0$ is valid for macrostate $\Psi$.

The H-theorem considers a dilute hard-sphere gas with no fine-structured energy or entropy barriers [2]. This model is ergodic and fully chaotic, which justifies $\mathcal{B}$ (see the discussions in Sections 2.7 and 5.1); therefore, the phase-space analysis in Section 2.5 readily establishes the second law. The simplified approach of Figure 2 can also be employed, based on the rigorous proof of the Boltzmann equation [21], without following the traditional route through the H-function: if the macrostate governing equations (the Navier-Stokes-Fourier equations) can be derived from the Boltzmann equation [21] and if their solutions are unique [22], then the second law may be obtained with the additional assumption of macrostate uniqueness.

### 3.2 Consistency of chaos

Another possible additional assumption is consistency of chaos. Under full chaoticity, a macrostate carries no detailed information about the specific dynamical processes of microstates.

It may be reasonable to assume that, in Figure 1(a), all macrostates in the same horizontal hyperplane exhibit a similar trend in $\dot{\Omega}$.

We can apply the slice-by-slice approach directly to Figure 1(a), without using the complicated phase-space subspace in Figure 1(b). Since $\dot{\Omega} \geq 0$ at the bottom of the funnel ($L_1$), regardless of whether $\partial\dot{\Omega}/\partial\Omega \leq 0$ or $\partial\dot{\Omega}/\partial\Omega > 0$, $\dot{\Omega} \geq 0$ is valid for every hyperplane. Thus, all macrostates satisfy $\dot{\Omega} \geq 0$.

**4. Boundaries of the second law of thermodynamics and beyond**

In essence, Inequality (7) ($\partial f/\partial\Omega > 0$) serves as the "driving force" that renders the otherwise non-preferential continuity equation of $f$ (Equation 1) directional. It is based on this fundamental assumption that $\dot{S} \geq 0$ is established. We now explore the boundaries of Inequality (7) and what may lie beyond, by examining an intuitive model.

4.1 Intrinsic-nonequilibrium steady state of Knudsen gas

It is well known that thermodynamics has its boundaries [e.g., 19]. Equilibrium is a global phenomenon arising from extensive particle-particle interactions [2] or the nonlocal nature of wave functions [14]. Without these mechanisms, a single classical particle, for example, is often governed by Newton's second law, not the second law of thermodynamics.

Perhaps the simplest many-particle non-thermodynamic system is a Knudsen gas [12], characterized by its particle mean free path ($\lambda_{\mathrm{F}}$) larger than the characteristic gas-body size $D$ (e.g., the gas-container size). Figure 3(a) depicts a Knudsen gas of $N$ particles (finite-sized hard spheres). The system is immersed in a thermal reservoir. The container is formed by thermal walls at temperature $T$. When a gas particle collides with a thermal wall, the reflected velocity is randomly sampled from the Maxwell-Boltzmann distribution at $T$ ($f_{\mathrm{MB}}$), uncorrelated with the incident velocity and without flux weighting. More analyses of the model are given in Appendix 1. The thermal-wall boundary condition is discussed in Appendix 2.

When $\lambda_F \gg D$, inside the container, the particle-particle collisions are sparse, and the particle trajectories are nonchaotic. Fast particles tend to rapidly collide with the thermal walls and release heat to the environment; slow particles tend to stay long in the interior. Consequently, in the steady state, the average particle kinetic energy $\bar{K}$ is smaller than its equilibrium value ($3k_BT/2$). Equivalently speaking, the effective gas-phase kinetic temperature, $T_k \triangleq 2\bar{K}/(3k_B)$, is lower than $T$.

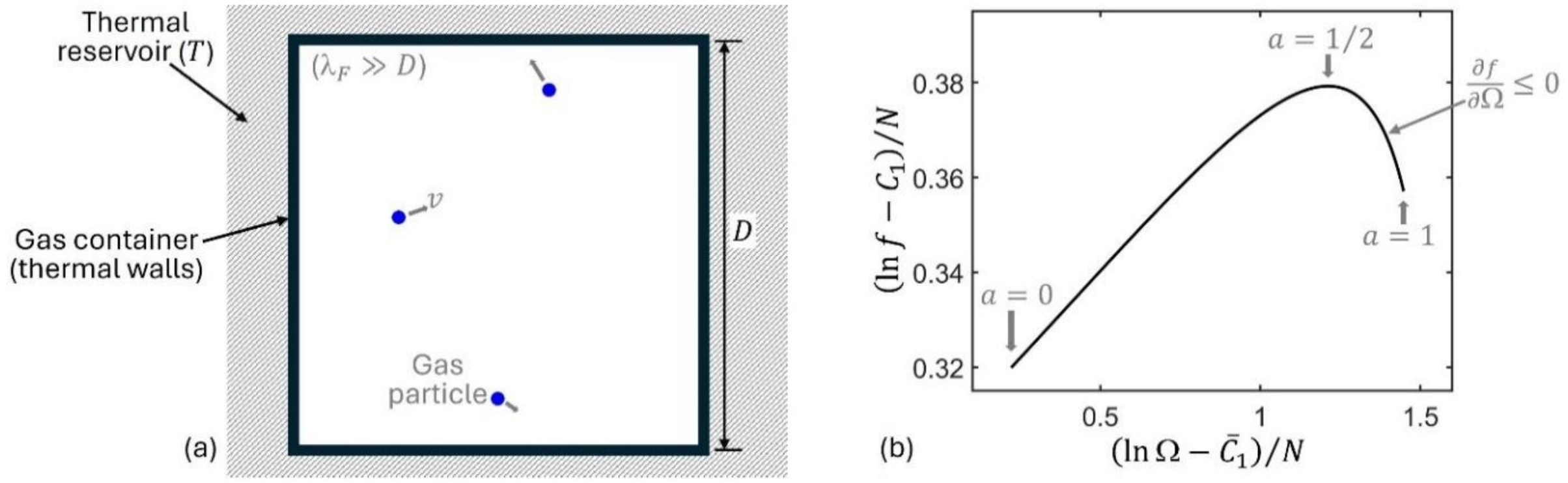

**Figure 3. (a)** A Knudsen-gas cell [12]. A number of gas particles (finite-sized hard spheres) randomly move in a gas container. The particle mean free path $\lambda_F$ is much larger than the container size $D$. The cell is immersed in a thermal reservoir and uses a thermal-wall boundary condition without flux weighting. Because particle-particle collisions are sparse, fast particles tend to collide rapidly with the container walls and release heat to the thermal reservoir, and slow particles tend to stay long in the interior. As a result, the internal energy of the gas phase remains below its equilibrium value, and the system cannot relax to thermal equilibrium. The probability of finding a particle of kinetic energy $K$ is inversely proportional to the particle speed $v$, leading to a non-Boltzmann $\rho$. This intrinsic-nonequilibrium state is unusual, because it is the steady state and is not sustained by any external thermodynamic driving force. **(b)** The $\ln f - \ln \Omega$ relationship of a set of macrostates of the Knudsen-gas model. Units are chosen such that $\beta = 1$. When $a = 1/2$, the cell walls are effectively in thermal equilibrium and, therefore, the macrostate probability ($f$) reaches the maximum value. Remarkably, $\partial f/\partial\Omega \leq 0$ from $a = 1/2$ to $a = 1$.

In other words, a Knudsen gas cannot relax to thermal equilibrium. For convenience, occasional particle-particle collisions are neglected. Under this simplification, the probability ($\rho_v$) of finding a particle with a momentum $\vec{p}$ is inversely proportional to the particle speed $v$, i.e., $\rho_v \propto 1/|\vec{p}|$. Accordingly, the probability density of microstate

$$\rho \propto \frac{e^{-\beta\epsilon}}{\kappa_p} \tag{10}$$

is non-Boltzmannian, where $\beta = 1/(k_BT)$, $\kappa_p = \prod_{j=1}^{N}|\vec{p}_j|$ is the kinetic factor, $\vec{p}_j$ is the momentum of the $j$-th particle, the microstate energy is $\epsilon = \sum_{j=1}^{N} K_j = \sum_{j=1}^{N}|\vec{p}_j|^2/(2m)$, and $K_j$ is the kinetic energy of the $j$-th particle. With a given $\epsilon$, various combinations of $|\vec{p}_j|$ can result in different $\rho$. Thus, $\mathcal{B}$ (Equation 2) is not satisfied.

The counterintuitive state of the Knudsen gas is referred to as "intrinsic nonequilibrium" to highlight its fundamental distinction from conventional nonequilibrium phenomena. An intrinsic-nonequilibrium state is a steady state, not a transient or fluctuating process. Moreover, the system is not driven away from equilibrium by any external thermodynamic force. The out-of-equilibrium state is rooted in the internal mechanism of nonchaoticity. Appendix 3 provides a more detailed discussion.

4.2 Negative correlation between $f$ and $\Omega$

To investigate the relationship between $f$ and $\Omega$ of the Knudsen gas, consider a set of macrostates characterized by $n_{\mathrm{w}}(K) = \xi_{\mathrm{a}} K^{a} e^{-\beta_{\mathrm{a}} K}$ $(0 \leq a \leq 1)$, where $n_{\mathrm{w}}(K)$ is the particle-number distribution with respect to kinetic energy $(K)$ immediately after collisions with the container walls; to keep $\int_0^{\infty} n_{\mathrm{w}}(\epsilon_{\mathrm{m}}) \mathrm{d}\epsilon_{\mathrm{m}} = N$ and $\int_0^{\infty} \epsilon_{\mathrm{m}} \cdot n_{\mathrm{w}}(K) \mathrm{d}K = 3Nk_{\mathrm{B}}T/2$, we set $\xi_{\mathrm{a}} = N\beta_{\mathrm{a}}^{a+1}/\Gamma(a+1)$ and $\beta_{\mathrm{a}} = 2(a+1)\beta/3$. When $a = 1/2$, $n_{\mathrm{w}}(K) \propto \sqrt{K} e^{-\beta K}$ is the Maxwell-Boltzmann distribution, i.e., the cell walls are effectively in thermal equilibrium. When $a \neq 1/2$, $n_{\mathrm{w}}(K)$ is non-Boltzmannian.

On the one hand, the macrostate probability $f$ is not proportional to the number of microstates $\Omega$. With different $\prod_{j=1}^{N} |\vec{p}_j|$, even if two microstates are at the same $\epsilon$, they may have different $\rho$ and, therefore, have different contributions to $f$. As the kinetic factor $\kappa_{\mathrm{p}}$ is taken into account, $f$ is dominated by the boundary condition, $n_{\mathrm{w}}(K)$. Similar to the calculation in [18], $f \propto \prod_i \left(g_i^{n_i}/n_i!\right)$, where $n_i$ is the number of particles at the $i$-th energy level $K_i$ after particle-wall collisions, and $g_i \propto 4\pi\sqrt{2K_i/m}$ is the density of states. Based on Stirling's approximation, $\ln f \approx \sum_i [n_i \ln(e g_i/n_i)] + C_0$, where $C_0$ is a system constant. The continuous form is obtained by replacing $n_i$ with $n_{\mathrm{w}}$:

$$\ln f = \int_0^{\infty} \left[\left(\xi_{\mathrm{a}} K^{a} e^{-\beta_{\mathrm{a}} K}\right) \cdot \ln\left(\frac{e_0}{\xi_{\mathrm{a}}} K^{1/2-a} e^{\beta_{\mathrm{a}} K}\right)\right] \mathrm{d}K + C_0 \qquad (11)$$

$$= N\left[a - \ln\frac{\beta_{\mathrm{a}}^{3/2}}{\Gamma(a+1)} + \left(\frac{1}{2} - a\right) \cdot \psi(a+1)\right] + C_1,$$

where $\psi(\bullet)$ denotes the digamma function, $C_1 = C_0 + N[\ln(e_0/N) + 1]$ is a system parameter independent of $a$, and $e_0$ collectively reflects Euler's number and the coefficient of $g_i$, including $4\pi\sqrt{2/m}$ and the configuration-space contribution.

On the other hand, the actual number of microstates $\Omega$ is determined not by the cell walls but by the macrostate configuration itself. Since $|\vec{p}| = \sqrt{2mK}$, the overall particle energy distribution $n(K) \propto n_{\mathrm{w}}(K)/\sqrt{K}$, i.e., $n(K) = \xi_{\mathrm{n}} K^{a-1/2} e^{-\beta_{\mathrm{a}} K}$, with the normalization factor being $\xi_{\mathrm{n}} = N\beta_{\mathrm{a}}^{a+1/2}/\Gamma(a+1/2)$. The internal energy is $U = \int_0^{\infty} [K \cdot n(K)] \mathrm{d}K = \kappa_{\mathrm{a}}(3Nk_{\mathrm{B}}T/2)$ and the effective gas-phase kinetic temperature is $T_{\mathrm{k}} \triangleq 2U/(3k_{\mathrm{B}}N) = \kappa_{\mathrm{a}} T$, where $\kappa_{\mathrm{a}} = (a+1/2)/(a+1)$. Similar to Equation (11),

$$\ln \Omega = \int_0^{\infty} \left[ \left( \xi_{\mathrm{n}} K^{a-1/2} e^{-\beta_{\mathrm{a}} K} \right) \cdot \ln \left( \frac{e_0}{\xi_{\mathrm{n}}} K^{1-a} e^{\beta_{\mathrm{a}} K} \right) \right] \mathrm{d}K + \bar{C}_0 \quad (12)$$

$$= N \left[ a - \ln \frac{\beta_{\mathrm{a}}^{3/2}}{\Gamma(a+1/2)} + (1-a) \cdot \psi\left(a + \frac{1}{2}\right) + 1 \right] + \bar{C}_1,$$

where $\bar{C}_0$ is a system constant and $\bar{C}_1 = \bar{C}_0 + N[\ln(e_0/N) - 1/2]$ is independent of $a$.

As $a$ is varied from 0 to 1, $(\ln f - C_1)/N$ and $(\ln \Omega - \bar{C}_1)/N$ are calculated from Equations (11) and (12), respectively (Figure 3b). Remarkably, in the high-$\Omega$ range, $\partial f/\partial \Omega \leq 0$, contradicting Inequality (7). When $a = 1/2$, since $n_{\mathrm{w}}(K)$ is in equilibrium, $f$ is maximized. When $a = 1$, $n(K)$ is apparently in the equilibrium form and $\Omega$ is maximized; yet, such a $n(K)$ corresponds to a non-Boltzmann $n_{\mathrm{w}}(K)$ and its probability ($f$) is lower than the maximum value. As $a$ changes from $1/2$ to 1, $\Omega$ keeps increasing while $f$ keeps decreasing.

### 4.3 Beyond the boundaries of the second law of thermodynamics

The nonchaoticity-induced $\partial f/\partial \Omega \leq 0$ is an inherent property of Knudsen gas and may lead to an unusual reduction in entropy. At the maximum $f$, $a = 1/2$ and the effective Knudsen-gas kinetic temperature is $T_{\mathrm{k}} = \kappa_{\mathrm{a}} T = 2T/3$ [12]. In the combined system of the gas phase and the thermal reservoir (Figure 3a), as the temperature field is nonuniform ($T_k < T$), the total entropy is lower than that of thermal equilibrium ($T_k = T$) (see Appendix Section 1.a). Therefore, although Knudsen gas is sometimes regarded as "unimportant," Knudsen-gas cells, when arranged in compound configurations (e.g., with locally nonchaotic energy or entropy barriers), can enable spontaneous entropy decrease at no energetic cost [12]. Such behavior is nontrivial: without any other effect, it allows for production of useful work by absorbing heat from a single thermal reservoir (Appendix Section 1.a) and spontaneous cold-to-hot heat transfer (Appendix Section 1.b) [11-14], thereby breaking the conventional boundaries of the second law.

Intrinsic nonequilibrium does not conflict with the principle of maximum entropy [11-13]. Under full chaoticity, subject to the constraints on $\rho_\alpha$ given by Equations (8) and (9), entropy $S =$ $-k_{\mathrm{B}}\sum_{\alpha}\rho_{\alpha}\ln\rho_{\alpha}$ is maximized through $\partial\mathcal{L}_{\mathrm{c}}/\partial\rho_{\alpha}=0$ [18], where the Lagrangian is $\mathcal{L}_{\mathrm{c}}=-k_{\mathrm{B}}\sum_{\alpha}\rho_{\alpha}\ln\rho_{\alpha}-\mu_{\mathrm{n}}(\sum_{\alpha}\rho_{\alpha}-1)$ for a microcanonical ensemble or $\mathcal{L}_{\mathrm{c}}=-k_{\mathrm{B}}\sum_{\alpha}\rho_{\alpha}\ln\rho_{\alpha}-\mu_{\mathrm{n}}(\sum_{\alpha}\rho_{\alpha}-1)-\beta_{\mathrm{n}}(\sum_{\alpha}\rho_{\alpha}\epsilon_{\alpha}-U)$ for a canonical ensemble, and $\mu_{\mathrm{n}}$ and $\beta_{\mathrm{n}}$ are the Lagrange multipliers associated with Equations (8) and (9), respectively. For a locally nonchaotic system, $\rho_{\alpha}$ may be directly affected by microscopic dynamics, e.g., Equation (10). It imposes additional dynamical constraints on $\rho_{\alpha}$, denoted by $C(\rho_{\alpha})=0$. The Lagrangian becomes $\mathcal{L}=\mathcal{L}_{\mathrm{c}}-\mu_{\mathrm{m}}C(\rho_{\alpha})$, where $\mu_{\mathrm{m}}$ is the additional Lagrange multiplier. Through $\partial\mathcal{L}/\partial\rho_{\alpha}=0$, entropy reaches a more constrained maximum [12], which is smaller than the global maximum at equilibrium. Regardless, $\partial\mathcal{L}/\partial\rho_{\alpha}=0$ ensures that entropy is maximized, compatible with the basic logic that the highest probability (measured by entropy) corresponds to the most probable state.

## 5. Extended discussion

### 5.1 The strong assumption: $\mathcal{B}$

In Section 2, the key to establishing the second law of thermodynamics is to assume $\mathcal{B}$ or, more generally, $\partial f/\partial\Omega>0$ (Inequality 7). Equation (3) suggests that $\mathcal{B}$ is a special case of $\partial f/\partial\Omega>0$. For example, consider a set of $\Omega_{\mathrm{t}}$ microstates; if $\rho_{\alpha}$ varies among microstates but its range of variation is less than $1/\Omega_{\mathrm{t}}$, $\mathcal{B}$ is not satisfied whereas $\partial f/\partial\Omega>0$ is ensured.

Ergodicity may justify $\mathcal{B}$ under appropriate conditions [17,23]. For a Hamiltonian system that is ergodic with respect to the invariant Liouville measure $q_{\mathrm{L}}$ on a common level set of the conserved quantities, if the equilibrium probability measure $P_{\mathrm{e}}$ is stationary and absolutely continuous with respect to $q_{\mathrm{L}}$, then $\mathrm{d}P_{\mathrm{e}}=\rho\cdot\mathrm{d}q_{\mathrm{L}}$. Liouville's theorem implies that $q_{\mathrm{L}}$ is invariant under the Hamiltonian flow, so that $\rho$ is invariant under the dynamics.

Full chaoticity may also justify $\mathcal{B}$, as discussed in Section 2.7. The present investigation, like Boltzmann's H-theorem [2], treats macroscopic behavior as an emergent phenomenon ultimately governed by the microscopic EOM. Accordingly, the probability of a microstate ($\rho_{\alpha}$) is determined by the parameters characterizing that microstate. As the extreme sensitivity to initial conditions renders specific microscopic information effectively inaccessible at the macroscopic level, $\rho_{\alpha}$ is assumed to be constrained only by normalization and the conserved quantities, which leads to $\mathcal{B}$. The absence of microscopic dynamics in the macrostate description is consistent with chaos theory [20]: despite deterministic governing equations, detailed microstate evolution is effectively unpredictable. Under energy conservation, the constraints on $\rho_\alpha$ are given by Equations (8) and (9), which constitute the standard formulation in statistical mechanics [18]: as the two equations constrain the maximization of entropy, the Maxwell-Boltzmann, Fermi-Dirac, and Bose-Einstein distributions are derived. In fact, in accordance with Section 2.5, Equations (8) and (9) themselves suffice to obtain these equilibrium distributions, since maximization of entropy (Inequality 6) follows from $\mathcal{B}$.

The general relationship between ergodicity and full chaoticity remains an important topic for future research. In Figure 1(a), the phase space is formed by all accessible microstates; regardless of ergodicity, inaccessible microstates do not affect the derivation of $\dot{\Omega} \geq 0$ at $\mathrm{L_y}$ in the phase-space subspace in Figure 1(b). For the dilute hard-sphere gas model considered in the H-theorem [2], ergodicity and full chaoticity are closely related. On the basis of the statistical independence of pre-collisional particle states ($\rho_{\mathrm{II}} = \rho_{\mathrm{a}}\rho_{\mathrm{b}}$) [21], it may be reasonable to take the $N$-body probability distribution as $\rho \approx \prod_{j=1}^{N} \hat{\rho}_j$, where $\hat{\rho}_j$ is the one-body probability distribution of the microstate of the $j$-th particle ($j = 1,2 \dots N$). Ergodicity and Liouville's theorem imply that all $\hat{\rho}_j$ are equal and, consequently, $\rho$ may be constant almost everywhere on the relevant invariant manifold in phase space.

### 5.2 The general assumption: Inequality (7)

At first glance, because $S = k_{\mathrm{B}} \ln \Omega$, $\partial f / \partial \Omega > 0$ (Inequality 7) seems to express that higher entropy is associated with higher probability. However, this interpretation of $\partial f / \partial \Omega > 0$ is over-simplified. In the present investigation, $f$ is defined as the probability in a fixed-width interval of multiplicity ($\delta\Omega$). In nonequilibrium statistical descriptions of isolated systems, $f$ can be explicitly time-dependent and is governed by probability flow among macrostates [15,24-27]. Therefore, $f$ may decrease even though typical phase-space evolution trajectories evolve toward macrostates of increasing $\Omega$ [e.g., 26,27]. In other words, the second law ($\dot{\Omega} \geq 0$ and $\dot{S} \geq 0$) does not necessarily imply $\dot{f} \geq 0$. The somewhat counterintuitive behavior $\dot{f} < 0$ can be attributed to the fact that the ensemble distribution changes over time: while an individual phase-space trajectory may move with $\dot{\Omega} \geq 0$, its probability can decrease if a growing fraction of the ensemble progresses to the larger-multiplicity regime.

As discussed in Section 2.6, according to Equation (1), $\dot{f} = \partial f/\partial t + \dot{\Omega}(\partial f/\partial \Omega) = -f(\partial\dot{\Omega}/\partial\Omega)$, where $\partial\dot{\Omega}/\partial\Omega$ is dominated by microscopic dynamics and can be either positive or negative. For instance, consider a one-dimensional $N$-particle ideal gas undergoing ballistic free expansion, with its length ($L_\mathrm{n}$) increasing at a constant rate $\dot{L}_\mathrm{n}$. With $\Omega \propto L_\mathrm{n}^N$ [2], it can be seen that $\dot{\Omega} \propto N\Omega\dot{L}_\mathrm{n}/L_\mathrm{n} > 0$ and $\partial\dot{\Omega}/\partial\Omega \propto (N-1)\dot{L}_\mathrm{n}/L_\mathrm{n} > 0$, such that $\dot{f} < 0$. The negative $\dot{f}$ may also be understood from the increasing Jacobian $J_\mathrm{F} \triangleq \mathrm{d}\Omega/\mathrm{d}L_\mathrm{n} \propto NL_\mathrm{n}^{N-1}$, which causes a fixed-width interval in $\Omega$ to correspond to a progressively narrower interval in $L_\mathrm{n}$.

Moreover, in the slice-by-slice analysis in Section 2, $\partial f/\partial\Omega > 0$ is required for every slice in the phase-space subspace in Figure 1(b) from $\mathrm{L_1}$ to $\mathrm{L_{y+1}}$. Except for $\mathrm{L_y}$ (macrostate $\Psi$), the other slices may not have a clear physical meaning. Without the assumption of macrostate uniqueness, these auxiliary slices can encompass portions of multiple macrostates without fully containing any of them, so that $f = \sum_\mathrm{S}\Omega$ may not clearly correspond to any particular macrostate and the definition of $S = k_\mathrm{B}\ln\Omega$ may not be relevant.

It is still not fully clear whether Inequality (7) is a sufficient and necessary condition for the second law, especially for locally nonchaotic systems. When $\dot{\Omega} = 0$, Equation (1) does not constrain $\partial f/\partial\Omega$; when $\dot{\Omega} > 0$, $\partial f/\partial\Omega > 0$ requests that $\partial(\ln f)/\partial t < -\partial\dot{\Omega}/\partial\Omega$. In Section 4 and Appendix 1, the Knudsen-gas models are examined based on their detailed dynamics, rather than solely through the calculation of $\partial f/\partial\Omega$. In Figure 3(a), $f$ is not governed by the macrostate configuration itself; what the macrostate configuration determines is $\Omega$.

When the system configuration is fully known, whether $\partial f/\partial\Omega$ is positive may be analyzed through the dynamical details, as shown in Section 4.2. In general, demonstrating $\partial f/\partial\Omega \leq 0$ does not require exploration of the entire phase space; examining a subset of macrostates is sufficient (e.g., Figure 3b). On the other hand, when certain details of the system are missing, the validity of $\partial f/\partial\Omega > 0$ may not be predetermined. In essence, statistical mechanics might not yield *a priori* knowledge about the probability space of unknowns that are, by definition, not known. In practice, empirical methods could be employed to test hypotheses against experiments.

5.3 Scope of the present investigation

It has long been understood that reversible microscopic dynamics can produce irreversible macroscopic behavior [e.g., 6,28]. For typical microstates of a chaotic system, entropy increases as the system evolves toward larger phase-space regions, with "typicality" being guaranteed by the large phase-space volume ratios. Velocity reversal is effectively unattainable due to instability under perturbations.

The present work advances this line of thinking, from a perspective distinct from that of the H-theorem. The H-theorem attempts to demonstrate that, in Newtonian mechanics, random particle-particle collisions drive the system toward equilibrium [2]. Effectively, the goal of the H-theorem is to analyze system evolution under Newton's laws over all possible initial microstates. However, as an emergent macroscopic phenomenon [20,28], the second law may not need Newton's laws (or the Schrödinger equation). For example, the classical ideal-gas law does not rely on Newton's second law [14]. As another example, suppose that, in a classical ideal gas in equilibrium, after every collision between two particles, their outgoing momenta are uncorrelated with their incident momenta and are randomly distributed according to the Maxwell-Boltzmann distribution; the equilibrium macrostate would remain unchanged, while momentum would no longer be conserved microscopically.

The derivation of $\dot{S} \geq 0$ in Section 2 does not involve any specific dynamical rule and, effectively, its goal is to analyze system evolution not only over all possible initial microstates but also under all possible equations of motion. In this sense, the phase-space analysis of Figure 1(b) is more general than the H-theorem. Equation (1) governs the macrostate-level probability $f$, and Inequality (5) reflects the statistical behavior of macrostates. They permit low-probability entropy-decreasing microstate evolution, compatible with fluctuations, Loschmidt's paradox, and the Poincaré recurrence theorem. Note that the details of fluctuations, such as their frequencies and magnitudes [e.g., 29], are beyond the scope of this manuscript.

Moreover, in contrast to the H-theorem, the time-irreversibility demonstrated by Inequalities (5) and (6) pertains to macrostate evolution, rather than microstate evolution. Because Inequality (6) is not based on microscopic EOM, the conclusion ($\dot{S} \geq 0$) is unrelated to the range of particle-particle interactions, the time scale, and the degree of nonequilibrium, and applies to both classical and quantum mechanics. For quantum systems, it is worth reiterating that our discussion is focused on the Boltzmann entropy $S$, which is not directly associated with the wave function ($\psi_{\mathrm{Q}}$). As long as the system is in a bound state with quantized energy and a nonlocal wave function, as the von Neumann entropy is zero, $\mathcal{B}$ is "trivially" satisfied and the second law is always obeyed [14]. The outcome of an observation, $|\psi_{\mathrm{Q}}|^2$, is probabilistic. Yet, $|\psi_{\mathrm{Q}}|^2$ is not $\rho$ but instead the consequence of quantum measurement, and its energetic effect and the role of entanglement [e.g., 30-33] are not part of the present study.

5.4 The "horizontal" microstate evolution

Figure 4 provides an intuitive illustration of a microstate evolution path in the funnel-shaped phase space. The increase in $\Omega$ along "height" can be related to the "horizontal" microstate evolution, i.e., evolution among the microstates that are in the same constant-$\Omega$ hyperplane. For example, in a classical $N$-particle ideal gas, small variations in particle positions within the interior of the gas body do not affect the effective gas-body volume ($V$) or the probability density distribution of particle momenta ($f_{\mathrm{T}}$), so that $\Omega = \Omega_{\mathrm{x}}(N, V, f_{\mathrm{T}})$ remains unchanged.

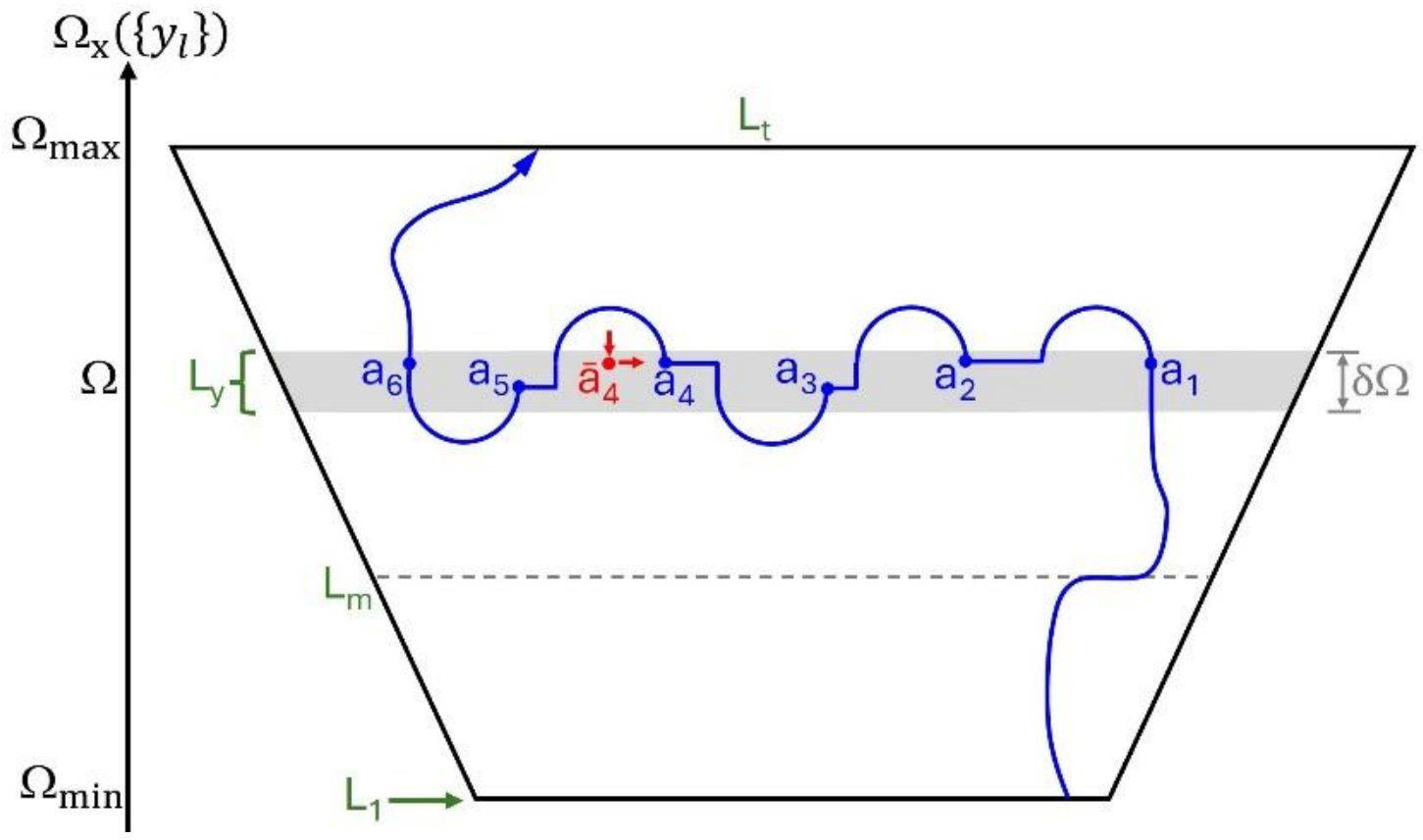

**Figure 4.** Schematic of a microstate evolution trajectory (the blue curve) in a funnel-shaped phase space. The band of $\mathrm{L_y}$ represents a constant-$\Omega$ hyperplane with a finite "layer" thickness ($\delta\Omega$); $\mathrm{a}_1 - \mathrm{a}_6$ are microstates along the trajectory, and $\bar{\mathrm{a}}_4$ is the time-reversed microstate of $\mathrm{a}_4$.

In the phase space, the "layer" size represents the number of microstates of the hyperplane. In hyperplane $\mathrm{L_y}$, the total horizontal segment of the evolution path (from microstate $\mathrm{a}_1$ to $\mathrm{a}_6$) is longer than those in lower hyperplanes (e.g., $\mathrm{L_m}$), allowing for a larger $\Omega$ at a higher "height." This mechanism is consistent with the fact that the microstates of a higher-entropy macrostate tend to be more likely to evolve into one another than those of a lower-entropy macrostate.

The horizontal (constant-$\Omega$) evolution of microstates is compatible with $\dot{\Omega} \geq 0$. The microstate evolution trajectory is unlikely to be perfectly smooth; rather, it may exhibit fluctuations, with the effects being amplified by the concentration of entropy at the hyper-interface in the ultrahigh-dimensional phase space. In $\mathrm{L_y}$, for instance, upon fluctuation, the trajectory may first move upward and then return to $\mathrm{L_y}$ from above (e.g., microstates $\mathrm{a_1}$, $\mathrm{a_2}$, and $\mathrm{a_4}$) or first move downward and then return to $\mathrm{L_y}$ from below (e.g., microstates $\mathrm{a_3}$ and $\mathrm{a_5}$). As the number of upward-evolving microstates exceeds that of downward-evolving ones, $\dot{\Omega}$ at $\mathrm{L_y}$ is positive. Since every upward-evolving microstate has a time-reversed counterpart that evolves downward (e.g., $\bar{\mathrm{a}}_4$, the time-reversed microstate of $\mathrm{a_4}$), the numbers of forward and reverse interface crossings between two adjacent constant-$\Omega$ hyperplanes are equal. However, the time-reversed trajectory also contains more upward-evolving than downward-evolving microstates, thereby preserving the same trend of $\dot{\Omega} \geq 0$ at $\mathrm{L_y}$.

Figure 4 merely shows an example of microstate evolution in a two-dimensional schematic and does not address the detailed upward and downward microstate evolution processes (e.g., dynamical mechanisms of fluctuations). More complex phenomena may exist in the actual ultrahigh-dimensional phase space. The horizontal evolution of microstates is not the direct reason for $\dot{\Omega} \geq 0$. The second law is derived from the assumption of $\mathcal{B}$ or $\partial f/\partial\Omega > 0$.

**6. Concluding Remarks**

To answer the question posed at the beginning of this paper: a micro-reversible system obeys the second law of thermodynamics ($\dot{S} \geq 0$) if all its components and processes (e.g., random-number generation algorithms) satisfy $\partial f/\partial\Omega > 0$ (Inequality 7), where $f$ is the macrostate-level probability in fixed-width intervals of multiplicity ($\Omega$). A sufficient but not necessary condition for $\partial f/\partial\Omega > 0$ is Boltzmann's assumption of equal *a priori* probabilities, denoted by $\mathcal{B}$. The requirement of $\partial f/\partial\Omega > 0$ implies that the second law may not be derived solely from the analysis of microstate-level probability (e.g., by using the Boltzmann equation), as macrostate properties ($f$ and $\Omega$) are taken into consideration.

Under the assumption of $\mathcal{B}$ or $\partial f/\partial\Omega > 0$, for an isolated system, the second law follows directly from the statistical and configurational properties of phase space (Equation 1) without relying on dynamical details, and is compatible with the possible entropy-decreasing evolution of individual microstates. The derivation is valid for both classical and quantum systems, regardless of the time scale, the degree of nonequilibrium, and the range of particle interactions. Additional assumptions (e.g., macrostate uniqueness or consistency of chaos) can reduce the complexity of the phase-space subspace.

In a fully chaotic system, detailed microscopic information is effectively inaccessible at the macroscopic level, and the microstate probability $\rho_\alpha$ is assumed to have no explicit dependence on the microscopic equations of motion (EOM). Consequently, $\mathcal{B}$ is justified, which in turn leads to the second law. Under appropriate conditions, ergodicity also justifies $\mathcal{B}$.

For the special fully chaotic and ergodic model of a dilute hard-sphere gas considered in the H-theorem, the second law may be established through the simplified approach in Section 3.1. This approach requires the Boltzmann equation but does not involve the traditional H-function argument.

Remarkably, for certain locally nonchaotic systems, as EOM explicitly influence $\rho_\alpha$, $\mathcal{B}$ may not be satisfied and $\partial f/\partial\Omega \leq 0$ is possible, so that the second law may not apply. A significant consequence is that entropy can decrease spontaneously at no energetic cost, allowing the system to reach an intrinsic-nonequilibrium steady state without an external driving force. It opens the possibility of spontaneous cold-to-hot heat transfer [12] and producing useful work by absorbing heat from a single thermal reservoir [11,34] with no other effect. Such a counterintuitive result does not challenge the second law within its established domain of application (fully chaotic systems) but instead extends the scope of thermodynamics beyond the conventional limits.

## Appendix

### 1. Intrinsic-nonequilibrium Knudsen-gas models

A Knudsen gas is defined as a rarefied gas in which the mean free path of the gas particles ($\lambda_F$) is larger than the characteristic gas-body size (e.g., the gas-container size $D$). As discussed for Figure 3(a), when a Knudsen gas is immersed in a thermal reservoir, it cannot relax to thermal equilibrium [12]. With the reflected velocities of gas particles at the container walls following the Maxwell-Boltzmann distribution $f_{MB}$ (see Appendix 2 for further discussion of the thermal-wall boundary condition), the effective kinetic temperature of the gas phase ($T_k$) is lower than the thermal-wall temperature ($T$). Such an unusual thermal effect has long been studied [e.g., 35,36]. However, in the past, the Knudsen-gas model was often regarded as "unimportant," because of the small system size. Another reason is that $T_k$ cannot be measured directly. If a wall-mounted temperature sensor is used to monitor the gas phase, the measured velocity distribution of the gas particles colliding with the sensor wall is $f_{MB}$, and the corresponding temperature is the environmental temperature $T$. The incoming and outgoing energy fluxes at the thermal walls are balanced, consistent with the zeroth law of thermodynamics.

The analyses in [12] (see Appendix Section 1.a) suggest that repeatedly converting a large ideal-gas body into a cluster of small Knudsen-gas cells (e.g., by inserting and withdrawing frictionless dividing walls) generates a cyclic heat flux between the gas and the surrounding thermal reservoir. The heat flux can drive a heat engine to produce useful work, incompatible with the conventional heat-engine statement of the second law.

Alternatively, when a Knudsen-gas cell is placed in a gravitational field [12] (see Appendix Section 1.b), since the descending particles accelerate while the ascending particles decelerate, the spatial distribution of $T_k$ is nonuniform. As a consequence, heat can be transported spontaneously by the gas particles from the cold side to the hot side, incompatible with the conventional refrigeration statement of the second law.

#### 1.a Knudsen-gas cell cluster

Figure 5(a) depicts the Knudsen-gas cell cluster studied in [12]. The left-hand side shows an ideal gas consisting of $N$ particles (finite-sized hard-spheres). The container is undivided, and the container size ($L_c$) is much larger than the particle mean free path ($\lambda_F$). Since the particle-particle collisions are extensive, the particle trajectories are chaotic. The system is immersed in a thermal reservoir at temperature $T$, and the container is formed by thermal walls. When a particle collides with a thermal wall, the reflected velocity randomly follows the Maxwell-Boltzmann distribution at the wall temperature ($T$), with no correlation with the incident velocity and without flux weighting (see Appendix 2 for more discussion). In equilibrium, the gas-phase kinetic temperature ($T_k \propto \bar{K}/k_B$) is $T$, where $\bar{K}$ is the average kinetic energy of the particles.

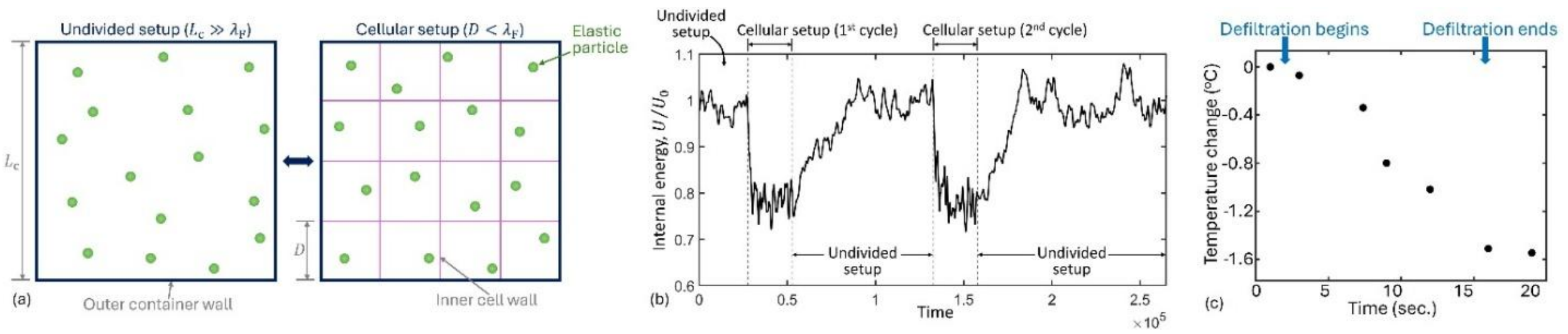

**Figure 5.** A Knudsen-gas model [12]. **(a)** An ideal gas (the undivided setup) is cyclically converted to a Knudsen-gas cell cluster (the cellular setup) by repeatedly inserting and withdrawing a set of frictionless inner cell walls. The container size ($L_c$) is much larger than the particle mean free path ($\lambda_F$); the cell size ($D$) is smaller than $\lambda_F$. In a Knudsen-gas cell, as $D < \lambda_F$, particle-particle collisions are sparse, i.e., the particle trajectories are nonchaotic. Fast particles tend to cross the cells rapidly and release heat to the cell walls, and slow particles tend to stay long in the interior. Consequently, the gas-phase internal energy ($U$) is lower than the corresponding equilibrium value ($U_0$). **(b)** Monte Carlo simulation result: a typical time profile of internal energy ($U$) in two cycles of cell-wall insertion and withdrawal. In each cycle, the first and second dashed vertical lines indicate cell-wall insertion and removal, respectively. **(c)** Experimental data of the temperature decrease when a nonwetting liquid defiltrates out of nanopores. Panels (b) and (c) are adapted from [12].

When a set of inner walls divides the large container into many small cells with the cell size $D < \lambda_F$, as shown on the right-hand side of Figure 5(a), each cell behaves as a Knudsen gas. The inner cell walls use the same thermal-wall boundary condition as the outer container walls. In a Knudsen-gas cell, particle-particle collisions are sparse. Compared with the equilibrium state, at any given instant, slow particles are more likely to be observed and fast particles less likely to be observed, so that the effective gas-phase kinetic temperature ($T_k$) is lower than $T$, reflecting the smaller gas-phase internal energy $U = \sum_{j=1}^{N} K_j \triangleq 3Nk_BT_k/2 < 3Nk_BT/2$.

With negligible particle-particle collisions, the energy distribution of the particles can be formulated as $n(\epsilon) \propto n_w(\epsilon)/\sqrt{\epsilon} = \xi_n \epsilon^{a-1/2} e^{-\beta_a \epsilon}$, and

$$T_k = \frac{2}{3k_B N}\int_0^\infty [\epsilon \cdot n(\epsilon)]d\epsilon = \frac{a+1/2}{a+1}T. \quad (13)$$

When the thermal walls are in thermal equilibrium, $a = 1/2$ and $T_{\mathrm{k}} \to 2T/3$ [12]. Only when $a$ is very large (i.e., the boundary condition is far from equilibrium), can $T_{\mathrm{k}} \to T$. Figure 5(b) shows a typical Monte Carlo (MC) simulation for the variation of $U$ during two cycles of cell-wall insertion and removal. The numerical results agree with the theoretical analysis.

In Figure 5(a), according to the first law of thermodynamics, when the undivided setup is changed to or from the cellular setup, the variation in $U$ is balanced by a heat exchange with the environment. As the inner cell walls are inserted and removed cyclically, the alternating heat flux can drive a heat engine to produce useful work by absorbing heat from the environment (a single thermal reservoir) without any other effect.

The inconsistency with the second law is rooted in the non-Boltzmann probability distribution of microstates ($\rho$). In the undivided setup in Figure 5(a), with extensive particle-particle collisions, $\rho \propto e^{-\beta\epsilon}$. In the cellular setup, as the particle trajectories tend to be nonchaotic, $\rho$ is non-Boltzmannian [12]:

$$\rho \propto \left(\prod_j \tau_j\right) \cdot e^{-\beta\epsilon}, \tag{14}$$

where $\tau_j \propto 1/v_j$ is the effective first-passage time of the $j$-th particle, and $v_j$ is the speed of the $j$-th particle. Because of $\prod_j \tau_j$, two microstates at the same energy level ($\epsilon$) may have different $\rho$, which does not satisfy $\mathcal{B}$ (Equation 2).

Another perspective on Figure 5 is to examine the spontaneous entropy decrease following the insertion of the inner cell walls. In the undivided setup, as the temperature field in the gas phase and the thermal reservoir is uniform, entropy reaches the equilibrium maximum $S_{\mathrm{eq}}$. With the inner walls, the effective temperature field becomes nonuniform ($T_{\mathrm{k}} \neq T$). As a result, $S$ decreases from $S_{\mathrm{eq}}$ to the constrained maximum $S_{\mathrm{ne}}$ [12]. Specifically, as thermal energy ($\Delta U$) is transferred from the gas particles to the thermal reservoir, the thermal-reservoir entropy increases by [18]

$$\Delta S_{\mathrm{R}} = \frac{\Delta U}{T}. \tag{15}$$

In Equation (15), the slight increase in thermal-reservoir temperature ($T$) is neglected as a conservative approximation; otherwise, the calculated $\Delta S_{\mathrm{R}}$ would be even smaller. The decrease in gas-phase entropy is

$$\Delta S_{\mathrm{G}} = -\int_0^{\Delta U} \frac{\bar{u}}{T_{\mathrm{k}}} \mathrm{d}\bar{u}. \tag{16}$$

Overall, because $T_{\mathrm{k}} < T$, $|\Delta S_{\mathrm{G}}| > \Delta S_{\mathrm{R}}$, and the total entropy variation $\Delta S_{\mathrm{R}} + \Delta S_{\mathrm{G}} < 0$.

The reduction in entropy from $S_{\mathrm{eq}}$ to $S_{\mathrm{ne}}$ is not driven by any external thermodynamic force; rather, it is caused by the inherent nonchaoticity. The cell-wall insertion is frictionless. Immediately after the cell walls are in place, the energy distribution of gas particles is initially in equilibrium and subsequently deviates from it, and this process is unforced. It occurs without an energetic cost and lies outside the scope of the conventional thermodynamic description.

Experimental studies have suggested that the kinetic temperature of a nonwetting liquid ($T_{\mathrm{k}}$) in nanopores may be lower than the ambient temperature ($T$) [12]. As the confined liquid defiltrated out of the nanopores, a significant temperature decrease was observed (Figure 5c). The change in thermal energy exceeded the total work dissipated by the nanofluidic flow. More comprehensive testing data on time-dependent locally nonchaotic entropy barriers were obtained using a nanoporous membrane treated on one side with bendable organic chains [11]. Across the membrane, the pressure of a dilute gas was spontaneously nonuniform.

## 1.b Knudsen gas in a gravitational field

Figure 6(a) depicts another Knudsen-gas model [12]. A low-height Knudsen-gas cell is in a uniform gravitational field ($g$). The top and bottom of the cell are two large horizontal thermal walls, between which the gas particles (finite-sized hard spheres) randomly move. The cell height ($z_0$) is much less than the nominal particle mean free path ($\lambda_{\mathrm{F}}$). The horizontal dimension is large. The left and right boundaries are subject to periodic boundary conditions.

If $z_0$ is relatively large (e.g., $z_0/\lambda_{\mathrm{F}} = 5$), particle-particle collisions are extensive, and the effective gas-phase kinetic energy ($T_{\mathrm{k}}$) is uniform along height $z$ (see Figure 6b). This phenomenon confirms that a column of chaotic ideal gas can reach thermal equilibrium in a gravitational field, as stated by Maxwell [37]: "…the temperature would be the same throughout (i.e., isothermal), or, in other words, gravity produces no effect in making the bottom of the column hotter or colder than the top."

If $z_0 \ll \lambda_{\mathrm{F}}$, (e.g., $z_0/\lambda_{\mathrm{F}} = 0.05$), particle-particle collisions rarely occur, and the particle trajectories tend to be independent of each other. All ascending particles become slower, and all descending particles become faster. As a result, even when the bottom-wall temperature ($T_{\mathrm{b}}$) is the same as the top-wall temperature ($T_{\mathrm{t}}$) (i.e., the system is immersed in a thermal reservoir), the expectation value of the particle kinetic energy ($\overline{K}$) is nonuniform along height $z$; that is, the system cannot relax to thermal equilibrium, as shown in Figure 6(b). Counterintuitively, $\bar{K}$ increases with $z$, i.e., the gas phase is effectively hotter at the top and cooler at the bottom, which should be attributed to the profile of the Maxwell-Boltzmann distribution function.

When $T_\mathrm{b}$ is mildly lower than $T_\mathrm{t}$, the gas particles can spontaneously transport heat from the cold side (the bottom thermal wall) to the hot side (the top thermal wall) without energetic penalty, as shown in Figure 6(c). It is incompatible with the conventional refrigeration statement of the second law of thermodynamics.

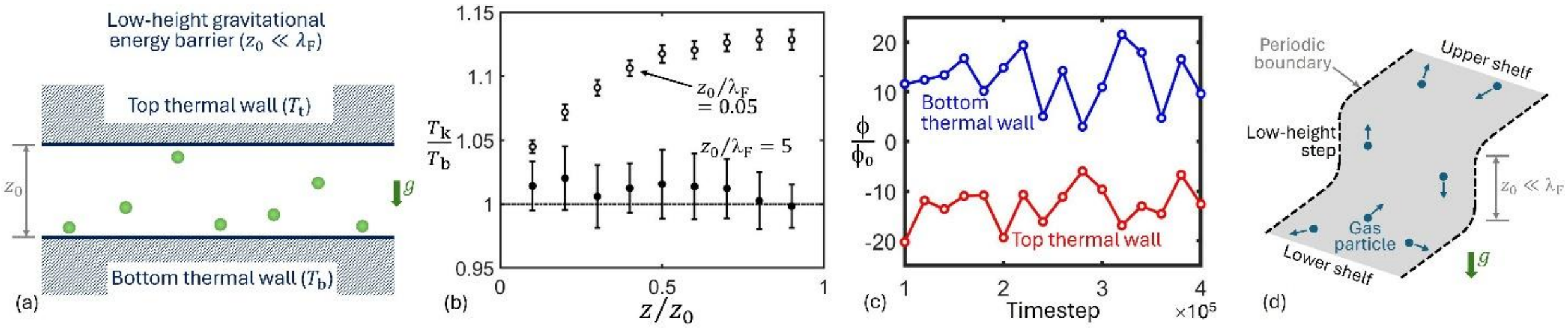

**Figure 6. (a)** A low-height Knudsen-gas cell in a uniform gravitational field ($g$). The cell height ($z_0$) is much smaller than the nominal particle mean free path ($\lambda_\mathrm{F}$). Since particle-particle collisions are sparse, the particle trajectories tend to be independent of each other. **(b)** Monte Carlo simulation results: Even when the bottom-wall temperature ($T_\mathrm{b}$) is the same as the top-wall temperature ($T_\mathrm{t}$) (i.e., the system is immersed in a thermal reservoir), the effective gas-phase kinetic temperature ($T_\mathrm{k}$) may be a function of height $z$, i.e., the system cannot relax to thermal equilibrium. When $z_0/\lambda_\mathrm{F}$ is large (i.e., the particle trajectories are chaotic), $T_\mathrm{k}$ is uniform and thermal equilibrium is reached. When $z_0/\lambda_\mathrm{F}$ is small (i.e., the particle trajectories are nonchaotic), $T_\mathrm{k}$ is highly nonuniform. **(c)** When $T_\mathrm{b}$ is mildly lower than $T_\mathrm{t}$, the gas particles can spontaneously transport heat from the cold side (the bottom wall) to the hot side (the top wall) without any energetic penalty. The curves show typical time profiles of the wall-to-gas heat transfer rates ($\phi$) at the top and bottom thermal walls, where $T_\mathrm{b}/T_\mathrm{t} = 0.95$, $z_0/\lambda_\mathrm{F} = 0.05$, and $\phi_0$ is the reference rate. A positive $\phi$ indicates that the gas particles absorb heat from the wall. **(d)** Monte Carlo simulation in [13], in which hard-sphere gas particles move randomly on the large horizontal upper and lower shelves connected by a low-height vertical step, in a uniform gravitational field ($g$). The step height $z_0 \ll \lambda_\mathrm{F}$. Particle motions are chaotic on the shelves and locally nonchaotic in the step. At steady state, across the low-height step, the particle-number-density ratio between the upper and lower shelves ($\delta$) is intrinsically non-Boltzmannian, similar to the effect of the $f_\mathrm{MB}$ boundary condition of thermal walls in panel (a) (Equation 17). Panels (b) and (c) are adapted from [12].

In Figure 6(a), with $z_0/\lambda_\mathrm{F} \ll 1$, in addition to the unusual thermal state, the particle number density distribution also cannot reach equilibrium [13]. When $T_\mathrm{t} = T_\mathrm{b} = T$, the steady-state particle-flux ratio ($\delta \triangleq n_\mathrm{t}/n_\mathrm{b}$) may be approximately assessed as [34]

$$\delta_1 \approx \int_{\sqrt{2gz_0}}^{\infty} p_\mathrm{z}(v_\mathrm{z})\mathrm{d}v_\mathrm{z} = 1 - \mathrm{erf}(\sqrt{\beta m g z_0}), \quad (17)$$

where $n_\mathrm{t}$ and $n_\mathrm{b}$ are respectively the numbers of particle-wall collisions at the top and bottom thermal walls, $v_\mathrm{z}$ is the vertical component of the particle velocity, and $p_\mathrm{z}(v_\mathrm{z}) =$ $\sqrt{2\beta m/\pi}\, e^{-\beta m v_z^2/2}$ is the one-dimensional Maxwell-Boltzmann distribution. In appropriate parameter ranges, $\delta_1$ is significantly smaller than the Boltzmann factor $\delta_0 = e^{-\beta m g z_0}$. More accurate results are obtained from Monte Carlo simulations [13], which confirm the trend $\delta_1 < \delta_0$. With no other effect, such an intrinsic-nonequilibrium phenomenon can spontaneously generate a macroscopic flow from microscopic thermal motions, and enable production of useful work in a cycle by absorbing heat from a single thermal reservoir.

In a recent experiment on supercapacitors [34], an intrinsic-nonequilibrium distribution of particle number density was observed. Compared to the "toy model" in Figure 6(a), the weak gravitational force was replaced by a strong Coulomb force, and the working medium was changed from a rarefied gas to a dilute electrolyte solution. The steady-state distribution of large ions in charged small nanopores was far from equilibrium, with no external driving force. The measured voltage variations were nearly one order of magnitude larger than the upper limit predicted by the second law of thermodynamics.

## 2. Boundary condition of the Knudsen-gas models

For the Knudsen-gas models in Figures 3, 5, and 6, to investigate the second law, the thermal-wall and isolated boundary conditions are both relevant, because neither provides an external driving force capable of taking the system away from equilibrium. Compared with the isolated case, the thermal-wall boundary condition is often more convenient to analyze.

In our research, at a thermal wall, the reflected velocities are fully randomized and sampled from the Maxwell-Boltzmann distribution ($f_{\mathrm{MB}}$) [12,13]. The $f_{\mathrm{MB}}$ boundary condition is employed under the assumption that the thermal wall has a large number of microscopic degrees of freedom associated with the thermal motion of its constituent atoms; upon collision with the wall, a gas particle exchanges energy and momentum with these microscopic degrees of freedom, and the interaction is sufficiently strong to eliminate memory of the incident velocity. Thus, the outgoing velocity is determined by the equilibrium statistics associated with the wall temperature $T$. The heat capacity of the wall is assumed to be much larger than that of the gas, so that the wall remains in a near-equilibrium state independently of the gas particles.

As demonstrated in the Monte Carlo simulations in [12], after gas particles are reflected from the container walls, chaotic collisions among the gas particles are responsible for establishing equilibrium in the gas phase. If the container size ($L_c$) is much larger than the particle mean free path ($\lambda_F$), the steady state is in thermal equilibrium, i.e., the gas-phase kinetic temperature ($T_k$) is close to the thermal-wall temperature $T$ (see the undivided-setup segments in Figure 5b). If the cell size $D < \lambda_F$, $T_k$ is significantly lower than $T$ (see the cellular-setup segments in Figure 5b). This mechanism of equilibration relies on particle-particle interactions, consistent with fundamental concepts of statistical mechanics [2]. With chaotic particle motion, the wall-to-wall return times (the time intervals between successive wall collisions of the same particle) are statistically equivalent for all particles and are much longer than $L_c/\sqrt{2\bar{K}/m}$. Regardless of whether the near-wall distribution of reflected velocities is proportional to $f_{MB}$, $v_z f_{MB}$, a uniform distribution, or any other distribution, the gas particles in the interior tend toward thermal equilibrium ($f_{MB}$). The bulk gas phase is largely insensitive to the boundary condition; that is, random particle-particle collisions can effectively "erase" information about the initial and boundary conditions.

Sometimes, in studies of a well-defined bulk gas phase, thermal-wall reflections are modeled using a flux-weighted distribution ($f_F$) [e.g., 38]. The $f_F$ boundary condition provides a convenient sampling algorithm that allows the thermal wall and the gas phase to attain thermal equilibrium without requiring collisions among the gas particles. It assumes that the re-emission of a gas particle from the wall is anisotropic with respect to the normal and tangential directions.

In the analyses of the small Knudsen-gas cells shown in Figures 3, 5, and 6, we do not use $f_F$ to describe the distribution of reflected velocities but instead adopt $f_{MB}$, for the following reasons: First, for the models in Figures 3 and 5, $f_F$ does not restore thermal equilibrium. With a bulk gas phase, $f_F$ typically assumes that the thermal wall is infinitely large. In contrast, in the interior of a small Knudsen-gas cell, although $f_F$ can ensure thermal equilibrium in the direction normal to the wall (associated with particle collisions with the opposite wall), the particle-velocity distribution in the lateral directions (associated with collisions with the side walls) is still inherently out of equilibrium, because $f_F$ reduces to the unweighted Maxwell-Boltzmann distribution ($f_{MB}$) in the tangential directions [38]. Therefore, the factor of $\prod_j \tau_j$ partially contributes to the microstate probability $\rho$ through transverse particle motions. The overall degree of intrinsic nonequilibrium would be decreased relative to that under the $f_{MB}$ boundary condition, but it would not vanish. Furthermore, even if only the velocity component normal to the thermal wall is considered, $f_F$ inherently results in a non-Boltzmann velocity distribution for particles colliding with the opposite wall. In Figure 3(a), with the $f_{\mathrm{F}}$ boundary condition, if a sensor is mounted on a cell wall, the detected gas-phase state would be far from equilibrium (based on $f_{\mathrm{F}}$), not the Maxwell-Boltzmann distribution ($f_{\mathrm{MB}}$)

Second, for the model in Figure 6, the Monte Carlo simulation in [13] (see Figure 6d) supports the $f_{\mathrm{MB}}$ boundary condition. In Figure 6(d), rarefied-gas particles (finite sized hard spheres) move randomly on two large horizontal shelves connected by a low-height vertical step ($z_0 \ll \lambda_{\mathrm{F}}$), in a uniform gravitational field $g$. The top and bottom edges of the step are open to the shelves, with no externally imposed boundary condition. The chaotic gas phase on the two shelves serves as a thermal reservoir, somewhat equivalent to the top and bottom thermal walls in Figure 6(a). Spontaneously, the upper-to-lower particle-number-density ratio between the two shelves ($\delta$) reaches an intrinsic-nonequilibrium steady state with a value close to $\delta_1$, as predicted by the $f_{\mathrm{MB}}$ boundary condition (Equation 17) and contrary to the prediction of the $f_{\mathrm{F}}$ boundary condition (the Boltzmann factor $\delta_0$). According to hard-sphere dynamics, when the pre-collisional particle-velocity distribution follows $f_{\mathrm{MB}}$ (as expected for the chaotic gas phase on the shelves in Figure 6d or for a thermal wall at equilibrium in Figure 6a), the post-collisional particle-velocity distribution also follows $f_{\mathrm{MB}}$. In other words, random particle-particle collisions naturally produce an effect similar to that of the $f_{\mathrm{MB}}$ boundary condition, and provide no mechanism for generating the non-Boltzmann distribution $f_{\mathrm{F}}$.

Third, a number of key assumptions underlying $f_{\mathrm{F}}$ may not hold. In a small Knudsen-gas cell, a well-defined bulk gas phase and a clear interfacial "gap" between the gas phase and the wall do not exist. This point is critical because such a gas-wall configuration is the basis for flux weighting. Moreover, the $f_{\mathrm{F}}$ boundary condition assumes that the wall is flat. If the wall contains narrow V-grooves (with the groove size much larger than the particle size), then, again because of the absence of a bulk gas phase, the direction associated with flux weighting is no longer uniquely defined by the overall wall normal but instead varies with the local groove-facet normal; the groove-facet orientations may be arbitrary.

At the fundamental level, for a bulk ideal gas, the $f_{\mathrm{F}}$ boundary condition may be reasonable because particle-particle collisions occur randomly in time. In a Knudsen gas, however particle-wall collisions dominate, and their occurrence is determined by the given container geometry and size. The fact that $f_{\mathrm{F}}$ does not require particle-particle collisions to establish thermal equilibrium (along the direction normal to the wall) is incompatible with the hypothesis of molecular chaos or full chaoticity. In a Knudsen-gas model, because there is no large gas phase with which the thermal wall is guaranteed to establish thermal equilibrium, thermal equilibrium should not be imposed *a priori*. Before a reflected gas particle can equilibrate with the surrounding gas, it may have already crossed the container and collided with another wall.

In fact, the specific form of the thermal-wall boundary condition is nonessential to the study of the second law. In the Monte Carlo simulations in [12,13], various boundary settings have been examined, with the degree of randomness of reflected velocities and the level of correlation with incident velocities being varied over broad ranges. As long as particle-particle collisions are sparse, no reasonable cell-wall boundary condition can maintain the Knudsen gas in equilibrium.

## 3. Intrinsic-nonequilibrium steady state

The unusual steady states in Figures 3, 5, and 6 are referred to as "intrinsic nonequilibrium" to emphasize the fundamental distinction from conventional nonequilibrium processes. First, the states are steady, not transient or fluctuating. Even if the system is initially in equilibrium, it would spontaneously evolve to the out-of-equilibrium steady state. Second, such a state is not sustained by any external thermodynamic force, e.g., a temperature or pressure gradient maintained by the environment. The models are isolated or immersed in a thermal reservoir, except for Figure 6(c).

Whether an intrinsic-nonequilibrium state can exist is a long-standing problem in statistical physics. For example, both Maxwell and Boltzmann studied the chaotic counterpart of the model in Figure 6(a): a column of ideal gas in a gravitational field [37]. According to the second law of thermodynamics, the gas-phase kinetic temperature must be independent of height at steady state. Otherwise, by using two ideal gases of different particle masses to form a double-column structure, a temperature difference can be generated spontaneously, enabling a heat engine to produce useful work by absorbing heat from a single thermal reservoir with no other effect. It has been known that this argument breaks down when the gas phase is "thin," i.e., when its thickness is smaller than the particle mean free path, in which case the second law may no longer be applicable [39].

Intrinsic nonequilibrium is compatible with the principle of maximum entropy [12]. As the locally nonchaotic particle movements are less random than those in a fully chaotic system, more information about the microstate probability ($\rho_\alpha$) may be known, i.e., $\rho_\alpha$ is more constrained than that of a chaotic ideal gas. The additional constraints arise from microscopic dynamics. As entropy is maximized, it reaches the more constrained maximum ($S_{\mathrm{ne}}$), which is smaller than the less constrained equilibrium maximum ($S_{\mathrm{eq}}$). Regardless of the additional constraints on $S_{\mathrm{ne}}$, entropy maximization ($\partial\mathcal{L}/\partial\rho_\alpha = 0$) preserves the basic logic: the most probable macrostate is the one with the highest probability (measured by entropy).

Intrinsic nonequilibrium is compatible with energy conservation. Although the particle-velocity distribution is non-Boltzmannian in the interior of a Knudsen-gas cell, the net heat flux through all thermal walls is zero [12]. Note that detailed balance is not necessarily required at the cell walls, as demonstrated by the model in Figure 6(a-c).

Intrinsic nonequilibrium should not be confused with perpetual-motion machine of the second kind, such as Maxwell's demon or Feynman's ratchet [12]. A perpetual motion machine attempts to exploit equilibrium components and chaotic processes to generate a nonequilibrium result. In contrast, the Knudsen-gas models are based on locally nonchaotic mechanisms that are inherently out of equilibrium. Their behaviors are expected to lie beyond the conventional framework of thermodynamics.

## References

1. Kosmann-Schwarzbach Y, Schwarzbach BE (2010). *The Noether Theorems*. Springer
2. Kardar M (2007). *Statistical Physics of Particles*. Cambridge Univ. Press.
3. C. Cercignani, R. Illner, M. Pulvirenti. *The Mathematical Theory of Dilute Gases* (Springer Sci., 1994)
4. I. Ampatzoglou, J. K. Miller, N. Pavlović. A rigorous derivation of a Boltzmann system for a mixture of hard-sphere gases. *SIAM J. Math. Anal* **54**, 2320-2372 (2022).
5. Goldstein, S., Lebowitz, J. L., Tumulka, R., Zanghì, N. Gibbs and Boltzmann entropy in classical and quantum mechanics. In: V. Allori (ed), *Statistical Mechanics and Scientific Explanation* (World Sci. Publ., 2020)
6. Lebowitz, J. L. Boltzmann's entropy and time's arrow, *Phys. Today* **46**, 32-38 (1993).
7. M. L. Bellac, F. Mortessagne, G. G. Batrouni. *Equilibrium and Non-Equilibrium Statistical Thermodynamics* (Cambridge University Press, 2009)
8. Pierre Monmarché. Hypocoercive relaxation to equilibrium for some kinetic models. *Kinetic and Related Models* **7**, 341-360 (2014).
9. E. Dolera. Exponential convergence to equilibrium for solutions of the homogeneous Boltzmann equation for Maxwellian molecules. *Mathematics* **10**, 2347 (2022)
10. Goldstein, S. Boltzmann's Approach to Statistical Mechanics. In: Bricmont, J., Ghirardi, G., Dürr, D., Petruccione, F., Galavotti, M.C., Zanghi, N. (eds.), *Chance in Physics: Foundations and Perspectives* (Springer, Berlin, 2001)
11. Qiao Y, Shang Z, Kou R (2021). Molecular-sized outward-swinging gate: experiment and theoretical analysis of a locally nonchaotic barrier. *Phys. Rev. E* **104**, 064133. *https://doi.org/10.1103/PhysRevE.104.064133*
12. Qiao Y, Z. Shang (2024). Second law of thermodynamics: spontaneous cold-to-hot heat transfer in a nonchaotic medium. *Phys. Rev. E* **110**, 054113. *https://doi.org/10.1103/PhysRevE.110.054113*
13. Qiao Y, Shang Z (2024). On the second law of thermodynamics: A global flow spontaneously induced by a locally nonchaotic energy barrier. *Physica A* **647**, 129828. *https://doi.org/10.1016/j.physa.2024.129828*

14. Qiao Y (2025). Searching for quantum non-thermodynamic phenomena, *Phys. Rev. E* **111**, 054108. *https://doi.org/10.1103/PhysRevE.111.054108*
15. N.G. Van Kampen. *Stochastic Processes in Physics and Chemistry* (Elsevier, 2007).
16. Calkin MG (1996). *Lagrangian and Hamiltonian Mechanics*. World Scientific Publ.
17. B. V. Gnedenko, A. Ya. Khinchin. *An Elementary Introduction to the Theory of Probability, Edited by J. B. Roberts* (W.H. Freeman, 1961)
18. R. K. Pathria, P. D. Beale. *Statistical Mechanics* (Butterworth-Heinemann, 2011)
19. J. R. Dorfman. *An Introduction to Chaos in Nonequilibrium Statistical Mechanics* (Cambridge University Press, Cambridge, 1999)
20. D. P. Feldman (2019). *Chaos and Dynamical Systems* (Princeton Univ. Press).
21. Y. Deng, Z. Hani, X. Ma. Long time derivation of the Boltzmann equation from hard sphere dynamics. *Annals of Mathematics*, in press.
22. R. Temam. *Navier-Stokes Equations: Theory and Numerical Analysis* (AMS Chelsea Publishing, Providence, RI, 2001).
23. I. P. Cornfeld, S. V. Fomin, Y. G. Sinai. *Ergodic Theory* (Springer, 1982)
24. R. Kubo, M. Toda, N. Hashitsume, *Statistical Physics II: Nonequilibrium Statistical Mechanics* (Springer, 1991).
25. C. W. Gardiner. *Stochastic Methods* (Springer, 2009).
26. C. Godrèche, J. M. Luck. Nonequilibrium dynamics of urn models. *J. Phys.: Condens. Matter* **14**, 1601 (2002)
27. M. Baldovin, L. Caprini, A. Vulpiani. Irreversibility and typicality: a simple analytical result for the Ehrenfest model. *Physica A* **524**, 422-429 (2019).
28. J. Cardy. *Scaling and Renormalization in Statistical Physics* (Cambridge Univ. Press, 1996)
29. de Groot SR, Mazur P (2013). *Non-Equilibrium Thermodynamics*. Dover Publ.
30. D. Wallace. *The Emergent Multiverse: Quantum Theory according to the Everett Interpretation* (Oxford Univ. Press, 2012)
31. A. Cabello, M. Gu, O. Gühne, J.-Å. Larsson, K. Wiesner. Thermodynamical cost of some interpretations of quantum theory. *Phys. Rev. A* **94**, 052127 (2016)
32. S. Deffner, J. P. Paz. Quantum work and the thermodynamic cost of quantum measurements. *Phys. Rev. E* **94**, 010103 (2016)

33. C. L. Latune, C. Elouard. A thermodynamically consistent approach to the energy costs of quantum measurements. *Quantum* **9**, 1614 (2025)
34. Y. Qiao, M. Wang (2026). Beyond the boundaries of the second law of thermodynamics: intrinsic-nonequilibrium distribution of large ions in charged small nanopores. *Physica A* **702**, 132097.
35. T. Hondou. Equation of state in a small system: violation of an assumption of Maxwell's demon. *Europhys. Lett.* **80**, 50001 (2007)
36. J. Denur. Speed-dependent weighting of the Maxwellian distribution in rarefied gases: a second-law paradox? *Found. Phys.* **37**, 1685-1706 (2007).
37. Maxwell JC (1872). *Theory of Heat* (Longmans, Green, and Co., London, 1872, p. 300)
38. Y. Sone. *Molecular Gas Dynamics: Theory, Techniques, and Applications*. (Birkhäuser, Boston, 2007).
39. Santiago J, Visser M (2019). Tolman temperature gradients in a gravitational field. *Eur. J. Phys.* **40**, 025604